\documentclass[preprint,12pt]{elsarticle}
\usepackage{epsfig}
\usepackage{color}
\usepackage{subfigure}
\usepackage{cprotect}
\usepackage{hyperref}
\usepackage[absolute]{textpos}
\newcommand{\CompHEP}{{CompHEP}}
\newcommand{\CalcHEP}{{CalcHEP}}

\def\GRACE{{\it GRACE}}
\def\FeynArt{{\it FeynArts/FormCalc}}
\def\HELAS{{\it HELAS}}
\def\MADGRAPH{{\it MADGRAPH}}
\def \C {{\it C}}

\def\heshe{he/she}
\def\hisher{his/her}

\def\REDUCE{{\it Reduce}}

\newcommand{\chvers}{{3.4}} 
\newcommand{\choldvers}{{2.3 }} 
\usepackage{setspace}
\begin{document}

\title{CalcHEP \chvers~for collider physics within and beyond the Standard Model}

\author{Alexander Belyaev$^{1,2}$, Neil D. Christensen$^3$, Alexander
  Pukhov$^4$\\\vspace{0.2in}
$^1$School of Physics \& Astronomy, University of Southampton,\\
Highfield, Southampton SO17 1BJ, UK\\
$^2$Particle Physics Department, Rutherford Appleton Laboratory,\\
Chilton, Didcot, Oxon OX11 0QX, UK,\\
E-mail: a.belyaev@phys.soton.ac.uk
\\
$^3$Pittsburgh Particle physics, Astrophysics  and Cosmology Center\\
Department of Physics \& Astronomy, University of Pittsburgh,\\
3941 O'Hara St., Pittsburgh, PA 15260, USA,\\
E-mail: neilc@pitt.edu\\
$^4$Skobeltsyn institute of Nuclear Physics of Lomonosov,\\
 Moscow State University, Russia,\\
E-mail: pukhov@lapp.in2p3.fr
}

\date{\today}

\begin{abstract}
We present version \chvers~of the \CalcHEP~software  package which is 
designed for effective evaluation and simulation 
of high energy physics collider processes at parton level.

The main features of \CalcHEP~are the computation   of  Feynman
diagrams, integration over   multi-particle phase space and event
simulation at parton level. The principle attractive key-points 
along these lines are that it has:
a) an easy startup even for those who are not familiar with \CalcHEP;
b) a friendly and convenient graphical user interface (GUI);
c) the option for a user to easily modify a model or introduce
   a new model by either using the graphical interface or 
   by using an external
   package with the possibility of cross checking the results
   in different gauges;
d) a batch interface which allows to perform very complicated 
   and tedious calculations connecting production and decay modes 
   for processes with many particles in the final state.
   
With this features set, \CalcHEP~can 
efficiently perform calculations with a high level of automation
from a theory in the form of a Lagrangian down to 
phenomenology in the form  of cross sections, 
parton level event simulation and various kinematical distributions.

In this paper we report on the
new features of \CalcHEP~\chvers~which improves the power of our package
to be an effective tool for the study of  modern collider phenomenology.

\end{abstract}

\begin{textblock*}{30ex}(\textwidth,5ex)
PITT PACC 1209
\end{textblock*}


\maketitle
\newpage

{\bf PROGRAM SUMMARY/NEW VERSION PROGRAM SUMMARY}

\begin{small}
\noindent
{\em Manuscript Title:}  CalcHEP \chvers~for collider physics within and beyond the Standard Model                                     \\
{\em Authors:}  Alexander Belyaev, Neil D. Christensen, Alexander Pukhov                                              \\
{\em Program Title:} CalcHEP                                         \\
{\em Journal Reference:}                                      \\
{\em Catalogue identifier:}                                   \\
{\em Licensing provisions:}  none                                 \\
{\em Programming language:}  C                                 \\
{\em Computer:} PC, MAC, Unix Workstations                                            \\
{\em Operating system:}  Unix                     \\
{\em RAM:} Depends on process under study                                              \\
{\em Number of processors used:} 1 for the graphical interface; 
up to the number available in batch mode                              
{\em Classification:} 4.4 Feynman diagrams, 5 Computer Algebra                                      \\
{\em External routines/libraries:} X11                           \\
{\em Nature of problem:}
1. Implement new models of particle interactions.
2. Generate Feynman diagrams for a physical process in any
implemented theoretical model.
3. Integrate phase space for Feynman diagrams to obtain cross sections
or particle widths taking into account kinematical cuts.
4. Generate unweighted events to simulate collisions at a modern collider.
   \\
{\em Solution method:}
1. Symbolic calculations.
2. Squared Feynman diagram approach
3. Vegas Monte Carlo algorithm.
\\
{\em Restrictions:} Up to $2\to 4$ production ($1\to 5$ decay) processes are realistic on
typical computers.  Higher multiplicities sometimes possible
for specific  $2\to 5$ and $2\to 6$  processes.
   \\
{\em Unusual features:}  Graphical user interface, symbolic algebra
calculation of squared matrix element, parallelization on a pbs cluster.
   \\
{\em Running time:} Depends strongly on the process.  For a typical
$2\to2$ process it takes seconds, $2\to3$ processes 
the typical runninc time is of the order of minutes.
For higher multiplicities it could  take much longer.
   \\
\end{small}
\newpage
\tableofcontents
\newpage


\section{Introduction}
{\bf \CalcHEP} ({\bf Calc}ulations in {\bf H}igh {\bf E}nergy {\bf
P}hysics)  is a package for the automatic  evaluation of
production cross sections and decay widths in elementary particle physics at the lowest order of perturbation theory
(i.e. in the born approximation) within various theoretical models of particle
physics including effective models.

\CalcHEP~is the next step in the development of the \CompHEP\cite{Pukhov:1999gg} package 
which was created by one of us (AP) together with  his colleagues from the
Skobeltsyn Institute of Nuclear Physics. 
The main idea of  \CalcHEP~is to provide
an interactive environment where the user can pass from the Lagrangian to the final distributions
effectively with a high level of automation.  
Other packages  created to solve  similar problems   are  
\GRACE\cite{GRACE,Yuasa:1999rg,Belanger:2003sd},
\HELAS\cite{HELAS}, 
\CompHEP{}\cite{Pukhov:1999gg,Boos:2004kh},
\FeynArt\cite{FeynArt,Hahn:2000kx,Hahn:2000jm},
\MADGRAPH\cite{Maltoni:2002qb,Alwall:2011uj}, 
HELAC-PHEGAS\cite{Kanaki:2000ey, Papadopoulos:2000tt,Cafarella:2007pc}, 
O'MEGA\cite{Moretti:2001zz}, WHIZARD\cite{Kilian:2007gr},
and
SHERPA\cite{Gleisberg:2003xi,Gleisberg:2008ta}.
Since the last published version \choldvers\cite{Pukhov:2004ca},
\CalcHEP~has been significantly improved
to become  an efficient and powerful tool for modern collider physics studies.

One of the main advantages of \CalcHEP~is a convenient interactive menu-driven 
{\bf G}raphical  {\bf U}ser {\bf I}nterface (GUI)
with detailed contextual help 
which can be viewed by pressing the \verb|F1| key.  
Also, the notations used in \CalcHEP~are
very similar to those used in particles physics.  These features, and
others, allow a beginner to start using \CalcHEP~right away even if
\heshe~has no prior experience with \CalcHEP.

Among the other important advantages of \CalcHEP~are: the ability to
create and/or modify models using either the 
internal graphical editor or by using external editors/packages,  the
option to use either Feynman or unitary gauge in the evaluation of
Feynman diagrams which provides a powerful cross check
of the model implementation and the numerical results, 
the ability to dynamically and automatically calculate the width of
unstable particles when the parameters of a model are changed, and the
ability to easily choose Feynman diagrams and squared Feynman diagrams
to remove from the calculation for a study of interference effects
(for example).

The \CalcHEP~package consists of two modules: a symbolic and a
numerical module. The symbolic session allows the user to dynamically work with a
physics model, symbolically calculate squared matrix elements, 
export the results as C-code and
compile the C-code into the executable \verb|n_calchep|. The
numerical module performs the evaluation of the integral over phase
space to determine the cross section or decay width of
the user-defined process.  It can also histogram the events to produce various kinematical
distributions taking into account user-defined kinematical
cuts.  Additionally, \CalcHEP~can be run in non-interactive mode
using various scripts provided for the user including the batch
interface summarized below.

Among the new  important features is a batch interface, which takes
the user's input and automatizes the calculation of the production and decay
processes and combines the results, connecting
production processes with decays, to produce a final event file in Les
Houches Event (LHE) format.  The final LHE file can be used in other
Monte Carlo (MC) software, including the MC software of the LHC
experiments.
The batch interface also supports
scanning over multiple parameters and parallelizes the entire calculation
over the processors of a multi-core machine or over the processors of
a high performance computing cluster which enables the use of hundreds
or thousands of processors for the calculation.  This last feature is
responsible for a significant enhancement in the speed of the symbolic
and numerical calculations.

The flexibility of \CalcHEP~allows to work with a variety of  Models Beyond the Standard Model
(BSM).
While the Standard Model  is included in the \CalcHEP~distribution,
various BSM models, for example, the SUSY Models MSSM, NMSSM, CPVMSSM \cite{Belyaev:1997jn,Belanger:2001fz, Belanger:2005kh, Belanger:2006qa},
the complete model with Leptoquarks~\cite{Belyaev:2005ew},
Little Higgs Model \cite{Belyaev:2006zz}, TechniColor Model \cite{Belyaev:2008yj}, MUED model \cite{Belanger:2010yx} and many others
in the \CalcHEP~format  are available to be  downloaded, imported  and used by \CalcHEP. 
The complete set of models for \CalcHEP~is available at the High Energy Physics Model Database (HEPMDB)~\cite{Leshouches2011} described in section \ref{HEPMDB}
and listed in Table~\ref{hepmdb-models}.

For SUSY models   there are external programs Isajet,
SuSpect, SoftSusy \cite{Allanach:2001kg},  SPheno \cite{Porod:2003um}, NMSSMTools
\cite{Ellwanger:2006rn}, CPsuperH \cite{Lee:2003nta}, SUSEFLAV
\cite{Chowdhury:2011zr}   for 
spectra calculation at  loop level. An interface with these programs is 
implemented for the SUSY models mentioned above and can be easily 
realized via the SLHAplus  package \cite{Belanger:2010st} included in \CalcHEP~for any BSM model.

There are two  public codes  intended for the
generation of \CalcHEP~format model files using, as input, a short model
definition in terms of field multiplets, model parameters, and a Lagrangian. They are LanHEP
\cite{Semenov:2008jy},  FeynRules \cite{Christensen:2008py} and SARAH \cite{Staub:2008uz}. 
Most of the \CalcHEP~models were generated by LanHEP. 

\CalcHEP~ can be used as a generator of matrix elements for external programs.
This option was realized in the micrOMEGAs \cite{Belanger:2006is,
Belanger:2010gh} package for calculation of Dark Matter  observables.
Development of \CalcHEP~was strongly influenced by the development of
micrOMEGAs. In section \ref{sec:getMEcode} we present tools which allow to
realize this option in user programs. In particular we present interface
between  \CalcHEP~ and Root packages which allows to generate and evaluate 
matrix elements under Root environment.  

This paper has the following structure: In Section \ref{sec:
  installation}, we briefly describe the steps to install \CalcHEP. In Sections \ref{sec:
  symbolic} and \ref{sec: numerical} we discuss the interactive
symbolic and numerical sessions of \CalcHEP, respectively. 
In Section \ref{collecting}, we describe how to work with the results
of the numerical session.   
 In Sections
\ref{Batch} and \ref{sec:batch interface}, 
we introduce the new features of non-interactive sessions together
with the new
batch interface. In Section \ref{models} we present the  models of new
physics which are implemented
in \CalcHEP~and discuss the core methods for implementing new models.
In Section \ref{sec: tools}, we describe some external tools for
implementing new models and the HEPMDB repository for models.  In
Section \ref{sec:getMEcode}, we describe how to use the matrix element
code from \CalcHEP~with external code. 
In Section \ref{sec-benchmarks}, we describe some benchmarks for
testing the \CalcHEP~installation.
In Section \ref{sec:conclusions}, we conclude.

There are several  tutorials devoted to \CalcHEP usage in High Energy
Physics, for example, \cite{Kong:2012vg},
\begin{verbatim}
https://indico.cern.ch/conferenceDisplay.py?confId=189668 
\end{verbatim}
as well as
\begin{verbatim}
http://www.hep.phys.soton.ac.uk/~belyaev/webpage/hep_tools.html
\end{verbatim} 
within the PhD course which is being given annually by Alexander Belyaev
at University of Southampton.

\section{\label{sec: installation}Installation}

The \CalcHEP~source codes, a complete manual corresponding to the current version
and a variety of Beyond the Standard Model (BSM) implementations for 
\CalcHEP~are presented on  the \CalcHEP~web site
\footnote{Another resource for BSM implementations for \CalcHEP~is HEPMDB which
is described in Section \ref{sec: tools}.}:
\begin{center}
\verb|http://theory.sinp.msu.ru/~pukhov/calchep.html|.
\end{center}

\CalcHEP~is designed to run on an assortment of UNIX platforms. The current
version has been tested on Linux and Darwin. To install \CalcHEP, the
user should unpack the downloaded file\\

\verb|tar -xvzf  calchep_|\chvers.tgz \\

This will create the directory \verb|calchep_|\chvers~which we refer
to in this paper as 
\verb|$CALCHEP|. To compile \CalcHEP, the user should \verb|cd| into
the \verb|$CALCHEP| directory   and run  \\

\verb|gmake |   or  \verb| make| \\
{
If successful, the user should get the following message at the end of the compilation\\
\verb|         #  CalcHEP has compiled successfuly and can be started.|\\
}  
In case of a compilation problem, the user can try to find a solution
in the \CalcHEP~manual or
ask the authors for help by e-mail. 
Once the package is compiled, the user should create a working
directory where the calculations will be performed.
We will refer to this directory as \verb|$WORK| throughout this paper.
To install this directory\footnote{One also can use \$CALCHEP/work  as a working
directory.}, the user should run 
\begin{verbatim}
   ./mkWORKdir  <directory name>
\end{verbatim}
where \verb|mkWORKdir| can be found in  the \verb|$CALCHEP| directory.
This command creates the directory \verb|$WORK=<directory name>| along with the
subdirectories
\begin{center}
\verb|models/  tmp/  results/  bin/ |,
\end{center}
which are intended  for the  user's theoretical particle physics
models,  temporary files and numerical session files. 
\verb|$WORK| also  contains the  scripts:
\begin{center}
\verb|./calchep | and \verb| ./calchep_batch |,
\end{center}
which launch the \CalcHEP~GUI   and batch  sessions, respectively.

By default only the Standard model is distributed together with CalcHEP package.
Other models can be download independently. 
A large set of BSM models can be downloaded  directly  from the \CalcHEP~WEB page. A more complete set
of models  is stored
in the repository  \\
\verb|                 http://hepmdb.soton.ac.uk/|\\
which we  present in Section~\ref{HEPMDB}.

\section{\label{sec: symbolic}Interactive GUI symbolic session}
\begin{figure}[htb]
\begin{center}
\includegraphics[width=\linewidth]{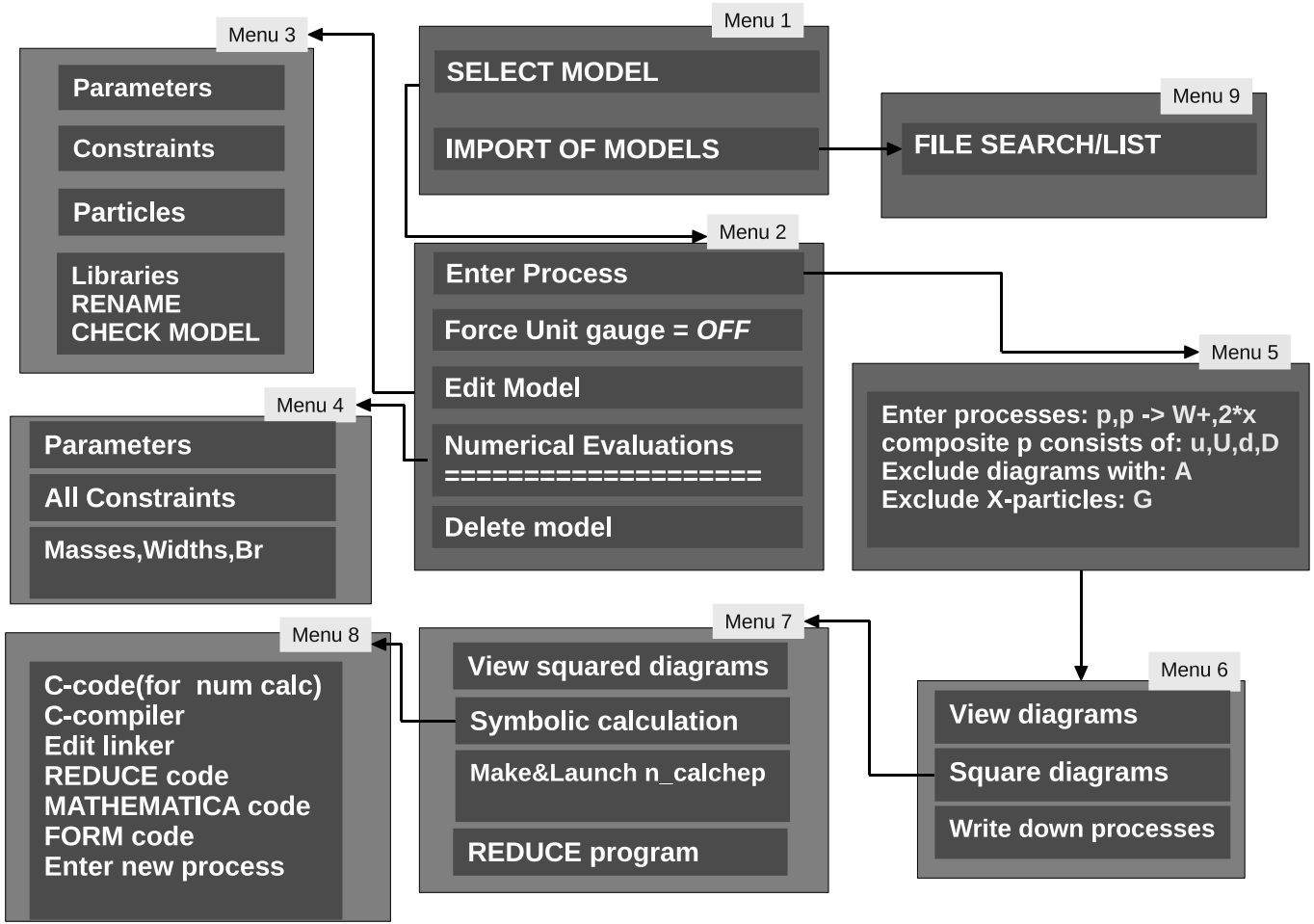}%
\caption{A menu-tree for the symbolic session of CalcHEP
 \label{symbolic_module}}
\end{center}
\end{figure}

\vskip 0.2cm  
\noindent
The menu structure for the symbolic session of \CalcHEP~is shown
schematically in 
Fig.~\ref{symbolic_module}.  These menus allow the user to:

\begin{itemize}

\item
 select a model  or import a new model from the file system [Menu~1].

\item implement  changes in a model [Menu 2,3] and check for syntax errors.


%

\item set a flag which forces  a calculation to be performed in
  physical (unitary)  gauge for   a model which has been  written in 
 t'Hooft-Feynman gauge [Menu 2];

\item check numerically the  mass spectrum, dependent parameter values and particle decay modes
before generating a process [Menu 4];
\item enter a scattering or decay process by specifying incoming and
  outgoing particles where 
$1 \rightarrow 2, \ldots ,1 \rightarrow 8$ decay processes
and 
 $2\rightarrow 1, \ldots , 2 \rightarrow 7$ scattering processes are
 supported [Menu 2,5];

\item   generate  Feynman diagrams,  display them, optionally exclude diagrams
as well as create the corresponding \LaTeX~output [Menu 6];
\item   generate, display and optionally exclude squared Feynman diagrams
as well as create the corresponding \LaTeX~output [Menu 7];
\item   calculate analytic expressions for squared matrix elements
      by using the fast built-in symbolic calculator [Menu 7];
\item   export the symbolic expressions of the squared diagrams
   in  REDUCE, MATHEMATICA or FORM format for  further
   symbolic and/or numerical  manipulations [Menu 7,8];
\item  generate optimized  C code for the squared matrix
    element  for further numerical calculations [Menu 8];

\item   launch the compilation of the generated C code and start the 
     corresponding numerical session [Menu 8];

\end{itemize}

Many new features have been added to the symbolic session compared to
version \choldvers \cite{Pukhov:2004ca}.  They are:
\begin{itemize}
\item[*] The option to evaluate the  $2\to1$ processes
is now available. It requires the 
distribution functions to be  used for the incoming particles
in the numerical session.

\item[*]  \CalcHEP~supports collision processes with  polarized massless
  fermions and vector bosons in the initial state.  
This is accomplished by ending
the particle's name with a \verb|%| as in:
\begin{center}
\verb|e%,A%->e,Z|
\end{center}

\item[*] \CalcHEP~now supports particles with spin up to 2 
(including  s= 0, 1/2, 1, 3/2 and 2).

\item[*] A new column has been added to the particle table allowing to enter
an ID for the particle from the  Monte Carlo numbering
scheme\cite{ParticlesFields}. This ID is used to interface the
model with 
parton structure functions, such as CTEQ and MRST,  and with Monte Carlo 
packages such as PYTHIA\cite{PYTHIA}. In other words, this ID
allows to interface a \CalcHEP~model with a package that uses
the Monte Carlo ID and is independent of the names given
to the particles.  \CalcHEP~does a basic sanity test on the Monte
Carlo ID's given to particles to make sure they do not clash with
the ID's of existing mesons and baryons from the PDG.  This helps
avoid potential problems with hadronization and fragmentation 
   when the \CalcHEP~events are being passed to external MC generators.

\item[*] CalcHEP includes the SLHAplus package \cite{Belanger:2010st} and
its functions can be used in CalcHEP model constraints.  It allows to 
\begin{itemize}
   \item[-] use an SLHA interface \cite{Skands:2003cj,Allanach:2008qq}  for external packages which calculate
particle spectrums,  mixing matrices of particle multiplets and  other model
constraints. SLHAplus allows to organize the corresponding interface
with minimal user programming;
   \item[-] implement tree level calculations for particle spectra and
mixing matrices for any  field multiplet; 
   \item[-] include QCD Yukawa corrections for Higgs-q-q vertices for
correct Higgs width calculations;
\end{itemize}
Further details can be found in Section \ref{sec:slhaplus}.

\item[*] In the current \CalcHEP~version,  we have added calculations
  for the Higgs-$\gamma$-$\gamma$ and Higgs-gluon-gluon effective
  couplings to the SLHAplus routines.  See section \ref{sec:HAA HGG}
  for further details.

\item[*] The width of a particle can now be calculated automatically during
a numerical evaluation when the parameters change.  To use this
feature, the user must add a \verb|!| to the beginning of the width
name in the particle table.  For example, to activate automatic calculation
of  the Higgs boson's width, the user should specify this in the
particle table in the following way 
\begin{verbatim}
   Full  name  |A  |A+ |PDG number |2*spin| mass |width |color|aux|
  Higgs        |h  |h  |25         |0     | Mh   |!wh   |1    |   |
\end{verbatim}
The exclamation mark  before the width symbol forces the width to be
calculated automatically.
\CalcHEP~does this by sequentially calculating the $1\to2$, $1\to3$, and $1\to4$ processes 
until a non-zero value for the width is obtained.  By default the  $1\to2$
decay processes are accompanied with $1\to3$ and  $1\to4$  ones with virtual W and
Z bosons. A proper matching with on-shell W/Z is done.  This is important for a correct calculation of the Higgs widths as well as for some BSM particles.
In  Section~\ref{sec:HAA HGG}, we present a comparision of branching fractions of Higgs
decays obtained with the Hdecay program and \CalcHEP. Calculation of channels
with  virtual W/Z can be switched off  via the \verb|F5| function key.

\item[*] \CalcHEP~has an option to use  particle widths calculated by an
external package.  If a \CalcHEP~model loads (in the \verb|Constraints| table) a file with a SLHA decay
table (see Sec.\ref{sec:slhaplus}), then  the automatic  width calculation
presented above uses the values read from the SLHA file instead of an internal matrix element 
calculation.  This behavior can be further controlled using an  option to read or
discard the widths from an SLHA file (see Section~\ref{sec:slhaplus}).
Also width parameters can get SLHA values using  SLHAplus functions
explisitely \cite{Belanger:2010st}.


\item[*]
Although SLHAplus contains many useful functions for model
implementation, we foresee the need to link external packages and user
written code to \CalcHEP.
This can be done in the new model table {\bf Libraries} [Menu 3] and 
greatly enhances
the facilities of \CalcHEP.  The user simply enters the path to the
compiled code to be linked when generating \verb|n_calchep|.
The Libraries table also  allows the user 
to define prototypes  of external
functions.
Two important uses for this functionality is to
link the LHAPDF\cite{Whalley:2005nh} structure function sets and user defined functions for phase space cuts 
and histograms.  See the details in Sections \ref{usrfun} and  \ref{extlib}.

\item[*] The \verb|Numerical Evaluations| menu [Menu 4] allows  
to evaluate the dependent parameters (constraints) as well as the
widths and branching ratios of the particles  before generating the
code for a specific process.  Like
the  automatic width  calculation, it uses the  dynamic linking facilities of modern
operation systems. During this procedure, the decay processes are
compiled  when they are needed   and stored
in the directory \verb|$WORK/results/aux| for subsequent usage  until the model is changed.  
\item[*]   \CalcHEP~can generate libraries of matrix elements to use with other
  packages (see Section \ref{sec:getMEcode}). These tools were first
  developed for the
micrOMEGAs package \cite{Belanger:2006is, Belanger:2010gh} to
dynamically generate and link  arbitrary matrix elements during runtime.
We also have supported  this option for  ROOT C++ interpreter.
\end{itemize}


\section{\label{sec: numerical}Interactive GUI numerical session}

\begin{figure}[htb]
\begin{center}
\includegraphics[width=\linewidth]{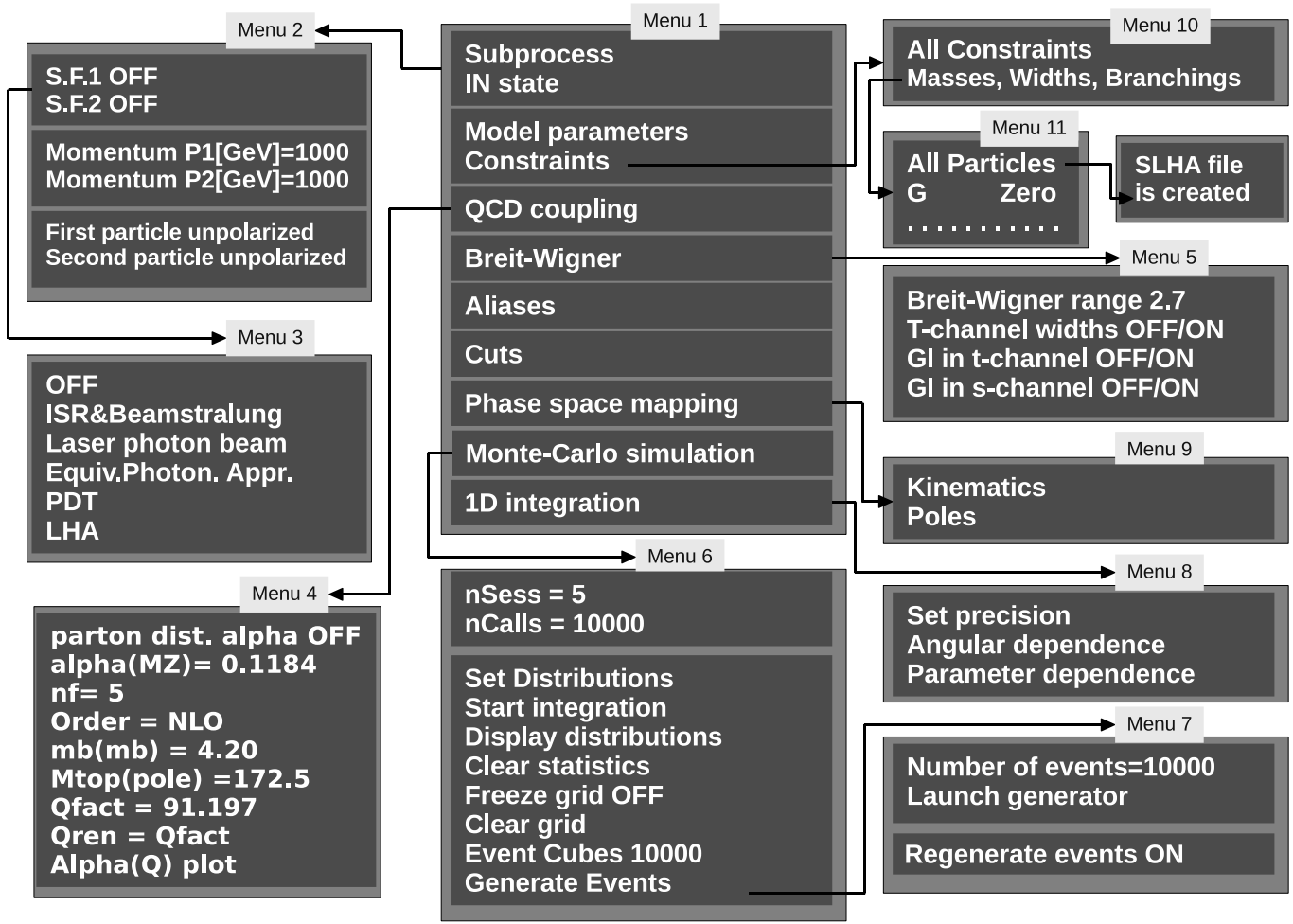}%
\caption{A menu-tree for the numerical session of CalcHEP
 \label{num_module}}
\end{center}
\end{figure}

\noindent
The menu system for the interactive   numerical session GUI of \CalcHEP~is schematically presented in
Fig.~\ref{num_module}. It  allows the user to:

\begin{itemize}

\item select a subprocess  for numerical processing if the generated
  code contains more than one;

\item set the momenta and helicities of incoming particles. The helicity is defined
with respect to the direction of motion.  For the  electron it is in
the interval
[$-\frac{1}{2}, \frac{1}{2}$].  A positive value (e.g. $\frac{1}{2}$)
corresponds to a right-handed  electron [Menu 1,2]; 

\item  convolute  the squared matrix element  with structure functions and
 beam spectra. 
\CalcHEP~comes with a set of parton distribution functions for protons
and anti-protons, initial state radiation and beamstrahlung spectra
for electrons, and the laser photon spectrum,  Weizsaecker-Williams
photon structure functions and proton photon structure functions for
the photon [Menu 3]. The user also has the option to use the LHAPDF
structure functions (see Sections \ref{sec:LHAPDF_linking} and \ref{extlib}   for details).
We note that the contents of this menu depend on the particle ID (see
Section \ref{particles}) and not on the particle's name;
   
\item   modify independent physical parameters such as the  coupling constants and masses which are
   involved in the process [Menu 1].  These parameters can be changed
one by one or the user can read the parameter values from a file using
this menu.  The file must contain one parameter on each line of the
file with the parameter name on the left and the numerical value on
the right separated by whitespace.  For example,
\begin{verbatim}
Mh  125
\end{verbatim}

\item view masses and dependent parameters (constraints) [Menu 1]. The list of
  dependent parameters which appear in this menu depends on whether
  they have been set to be ``public'' in the model files (see Section
  \ref{constraints});

\item calculate and view particle widths and decay fractions. This can
  be done via the \verb|Constraints| item of [Menu 1].  This menu also
  allows to write the full list of particle masses, widths and
  branching ratios to a file in SLHA format;

\item choose the scale used in the evaluation of the QCD coupling constant and
   parton structure functions  [Menu 4]. We also provide the
user the option to define the normalization and factorization 
scales independently. 

\item  apply various kinematical cuts [Menu 1 - Cuts option];

\item  define the kinematic scheme (phase space parametrization) in
  order to improve the efficiency of the 
Monte Carlo (MC)  integration and also to introduce a phase space
mapping in order to 
smooth the sharp peaks of a squared matrix element and structure functions [Menu 1 - Phase space mapping option];

\item   perform  a   Monte Carlo  integration of the phase space by VEGAS [Menu
1,6] (see Section \ref{MC});

\item   generate partonic level  events [Menu 1,6,7] (see Section \ref{MC});

\item  set and display distributions of various kinematical variables.
   It is also possible to export the distribution to file for use in
   external programs such as \LaTeX, gnuplot, PAW and Mathematica [Menu 6];

\end{itemize}

\subsection{\label{sec:LHAPDF_linking}
Built-in  and LHAPDF parton
  distribution functions}

\CalcHEP~comes preinstalled with several parton distribution function sets
These sets  and tools for  updating them have been  described in
\cite{Pukhov:2004ca}. To use one of these built-in PDFs, the user
should choose the PDT
({\bf P}article {\bf D}istribution {\bf T}ables)  item of Menu 3.   
A more comprehensive set of PDFs 
is available in the LHAPDF \cite{Whalley:2005nh}  library
which the user can install separately and link to a \CalcHEP~model by
using the \verb|Libraries| model table.
The source code for LHAPDF can be downloaded from the URL:
\begin{center}
\verb|http://projects.hepforge.org/lhapdf/|
\end{center}
which also contains instructions for the installation of LHAPDF.
To use the LHAPDF parton distribution functions in  \CalcHEP,
the user should add the line
\begin{verbatim}
-L<path_to_lhapdf>  -lLHAPDF
\end{verbatim}
to the  \verb|Libraries| model table.
This is sufficient to generate and compile the code for a process
which creates the executable  \verb|$WORK/results/n_calchep|.
However, it may not be sufficient to launch it.
If  \verb|libLHAPDF.so|  is located in a system area such as
\verb|/usr/lib|, then the library will be detected automatically.
Otherwise, information about the location of the shared library needs
to be provided with environment variables.
We recommend to add the  instructions
\begin{verbatim}
export  LD_RUN_PATH=<path_to_lhapdf>
\end{verbatim}
to the \verb|calchep| script in the \verb|$WORK| directory and
\begin{verbatim}
export  LD_LIBRARY_PATH=$LD_LIBRARY_PATH:<path_to_lhapdf>
\end{verbatim}
to the \verb|calchep_batch| script in the same directory where
\verb|<path_to_lhapdf>| is the path to the LHAPDF library. 
By setting \verb|LD_RUN_PATH| in the \verb|calchep| script,
  the user does not need to set this environment variable again before
  starting the generated
\verb|n_calchep| later. Also, using 
\verb|LD_RUN_PATH| in the \verb|calchep| script allows the LHAPDF
library to be updated without the user recompiling 
\verb|n_calchep|.
 If the linking of LHAPDF is  successful then  the \verb|LHA| item will  appear 
in Menu 3. 


\subsection{\label{sec:decaySLHA}
SLHA formatted files}

The {\it Constraints} menu allows the user to generate 
a complete SLHA file with all the particle's properties
for a  model including the
particles' masses, widths and branching ratios.
Since this file is is generate in SLHA format, the resulting file 
can be used by external programs such as MC generators.

To create this SLHA file,
the user should choose the \verb|Constraints| option of Menu 1,
go to the  \verb|Mass, Widths, Branching| submenu [Menu 10] and then choose the
\verb|All Particles| option [Menu 11].
When this option is chosen,
\CalcHEP~calculates the widths and branching ratios for the particles
and writes all the information in the SLHA file.
When it is finished, \CalcHEP~displays the meassage
\begin{center}
\verb|See results in file 'decaySLHA_n.txt'|
\end{center}
on the screen, where \verb|n| is an integer.
We note that these SLHA files can also be created during the symbolic
session through the \verb|All Constraints| menu item (Menu 4 of Fig. \ref{symbolic_module}).

It is well known that for a correct calculation of the Higgs width and
branching fractions, the QCD loop corrections must be included.  This
can be done by using running quark masses and yukawa couplings.  In
the default \CalcHEP~models, we include the following running quark masses which are
coded in the SLHAplus package
\begin{verbatim}
    McRun(Q)   MbRun(Q)   MtRun(Q) 
\end{verbatim} 
and the effective quark masses
\begin{verbatim}
    McEff(Q)   MbEff(Q)   MtEff(Q)  
\end{verbatim} 
which depend on the QCD scale (Q) and should be used to calculate the
Yukawa couplings (in the Constraints table).
The effective masses calculated
at the Higgs mass scale provide the correct partial Higgs width at
NNLO.  Parameters  with the scale \verb|Q| as an argument have a
special meaning in \CalcHEP.  When \CalcHEP~calculates the particle
width it substitutes the particle mass for the scale \verb|Q|.
Further details can be found in \cite{Belanger:2010st}.

To keep gauge invariance in tree level calculations, the particle's
pole mass must be the same as is used for the Yukawa coupling.  For
the {\it c} and {\it b} quarks, the effect is small at high energies.
However, for the {\it t} quark  the usage of the
effective mass can lead to the wrong decay modes for BSM
particles.

\subsection{\label{functions}Built-in kinematical functions}

\CalcHEP~ has a wide set of built-in
kinematical phase space functions which can be  used  to implement various kinematical
cuts and/or to construct distributions
\footnote{We note that the user can also write
  \hisher~own routines for kinematical functions.  Further details can
  be found in Section \ref{usrfun}.}. 
These functions are defined via the names of the 
outgoing particles and are called with the syntax
\begin{center}
 \verb| Name[^,_](P1[,P2,P3...])|,
\end{center}
where \verb|Name| is one capital letter specifying the function, the \verb|^| and \verb|_| 
are optional function modifiers useful in the case of identical outgoing
particles (described below), and \verb|P1[,P2,P3...]|
are the outgoing  particles involved in the observable.  
The available functions are:
\begin{itemize}
\item[A :] {\tt A(P1[,P2,...])} gives the angle between {\tt P1} and
  the combined momentum $p_{P2}+p_{P3}+\cdots$.  If only one particle is
  specified, as in {\tt A(P1)}, the angle between {\tt P1} and the
  first incoming particle is returned.  The angle is given in degrees.
\item[C :] {\tt C(P1[,P2,...])} gives the cosine of the angle defined
  above for {\tt A(P1[,P2,...])}.
\item[J :] {\tt J(P1,P2)} gives the jet cone angle between {\tt P1}
  and {\tt P2}.  The jet cone angle is defined as $\sqrt{\Delta y^2+\Delta\varphi^2}$, where  
$\Delta y$ is the difference in pseudo-rapidity and $\Delta \varphi$
is the difference in azimuthal angle between {\tt P1} and {\tt P2}.
\item[E :] {\tt E(P1[,P2,...])} gives the energy of the combined momentum
    $p_{P1}+p_{P2}+\cdots$.
\item[M :] {\tt M(P1[,P2,...])} gives the invariant mass of the
  combined momentum $p_{P1}+p_{P2}+\cdots$.
\item[P :] {\tt P(P1,P2[,P3,...])} first
  boosts into the cms frame of {\tt P1,P2[,P3,...]} and then takes the
  cosine of the angle between {\tt P1} (in the cms frame) and the boost direction.
\item[T :] {\tt T(P1[,P2,...])} gives the transverse momentum of the
  combined momentum $p_{P1}+p_{P2}+\cdots$.
\item[Y :] {\tt Y(P1[,P2,...])} gives the rapidity of the combined
  momentum $p_{P1}+p_{P2}+\cdots$.
\item[N :] {\tt N(P1[,P2,...])} gives the pseudo-rapidity of the
  combined momentum  $p_{P1}+p_{P2}+\cdots$.
\item[W :] {\tt W(P1[,P2,...])} gives the transverse mass of the
  particle set $S=\{${\tt P1,P2,...}$\}$ given by
$$\sqrt{ \left(\sum_{i \in S}
\sqrt{m_i^2+(p_i^T)^2}\right)^2 - \left(\sum_{i \in S}
\overrightarrow{p}_i^T\right)^2}$$
where $m_i$ and $p_i^T$ are the mass and transverse momentum
respectively of the $i$th particle. 
 


\end{itemize}
For example, \verb|M(m,M)| returns the invariant mass of a $\mu^-$, $\mu^+$
pair.

In situations where there is more than one identical outgoing particle,
all permutations of that particle are tested for cuts while their histograms are
added.  For example, suppose we have the process 
$p,p\to j,j,j$ where $j$ is a jet.  The cut \verb|J(j,j)| would be applied
to all six possible combinations of the three jets.
On the other hand, if this observable
were used for a distribution, each of the six combinations
would be binned for the histogram.  In other words, the histograms for
each combination are added together.  On the other hand, each particle is only used once
for each observable.  So, for example, \verb|J(j,j)| is never the jet
cone angle of a jet and itself.
The optional modifiers \verb|^|
and \verb|_| select the maximum and minimum value for the kinematical
function among all permutations.  For example, \verb|J_(j,j)|
calculates the jet cone angle of all pairs of jets and returns the
smallest one for the cut or histogram.
 
Although the momenta of incoming  particles can not be measured directly,
distributions of these momenta can still be interesting and useful to
understand the details of the collision.
For this, \verb|E1| and \verb|E2|can be used for the energy of the
first and second incoming particle, respectively.  The total partonic
CMS energy can be obtained with \verb|M12|.

\subsection{\label{sec: num alias}Aliases}
Aliases can be defined (see Menu 1 in
Figure \ref{num_module}) for particle sets which will be substituted in
the kinematical functions presented above.
The user should enter each alias on a
separate line where the name of the alias belongs in the first column and the
particles that are contained in the alias belong in the second
column.  For example, a jet could be defined as:
\begin{verbatim}
   Name | Comma separated list of particles 
   j    |u,U,d,D,c,C,s,S,b,B,G
\end{verbatim}
in the default SM.
A positively charged lepton of first or second generation might be
defined as:
\begin{verbatim}
   Name | Comma separated list of particles
   l+   |E,M
\end{verbatim}
in the default SM.  The user can make any alias definitions
\heshe~likes as long as the names do not match any particle names.
These aliases can be used in the cuts and distribution definitions.
For example, a $p_T$ cut can be placed on  QCD quarks and guons  with one
line using an alias (\verb|T(j)| rather than specifying a separate cut for each
quark and guon. 
Note that all particles included in an alias are treated  in the same manner as
identical particles. So, all combinations are tested for cuts and binned
for histograms.

\subsection{User defined kinematical  functions} \label{usrfun}

To implement a new  observable for cuts and distribution  one can 
write  a \verb|usrfun| routine which should be a C-code function.
The code of this function can be linked to the numerical session by
adding it to the \CalcHEP~model \verb|Libraries|
table. The user can then call these observables as \verb|Ucode| where
\verb|code| is a user-defined string which is passed to the
\verb|usrfun| routines for processing.  These user-defined observables
can be used in cuts and
distributions. The full prototype of the {\tt usrfun} function  is 
\begin{verbatim}
double usrfun(char *code, int nIn, int nOut, double *pvect, char **pName, int *pdg);
\end{verbatim}
where
\begin{itemize} 
   \item[-] \verb|nIn| is the number of incoming particles. 
   \item[-] \verb|nOut| is the number of outgoing particles.   
   \item[-] \verb|pvect| is an array containing the momenta of the incoming
     and outgoing particles.  The momentum of the $i^{th}$ particle (\verb|qi|)
     where $i$ goes from $0$ to $nIn+nOut-1$ is given by
\begin{center}
\verb|qi[k]=pvect[4*i+k]|
\end{center}
where \verb|k| goes from 0 to 3.  \verb|qi[0]=pvect[4*i]| gives the
energy (which is always positive), \verb|qi[3]=pvect[4*i+3]| gives the
component of the momentum along the axis of collision and \verb|qi[1]|
and \verb|qi[2]| give the transverse components of the momentum.
  \item[-]  \verb|pName| is an array of strings which contain the names of
    the particles.  \verb|pName[0]| gives the name of the first
    incoming particle, \verb|pName[1]| gives the name of the second
    incoming particle, \verb|pName[2]| gives the name of the first
    outgoing particle, and so on.
    \item[-] \verb|pdg| is an array of PDG ID codes for the
      particles.  
\end{itemize}
Some further functions which the user may find helpful are:
\begin{verbatim}
    int findval(char *name, double *value);
    int qnumbers(char *pname, int *spin2, int *charge3, int *cdim);
\end{verbatim}
The first accepts a parameter name (independent or dependent) and
fills the value of the parameter.
The second accepts a particle name and fills the quantum numbers of
the particle.  Further details of these functions can be found in
Section \ref{sec:getMEcode}.

We provide an example \verb|usrfun| routine in the file
\verb|$CALCHEP/utile/usrfun.c| 
which calculates the  transverse mass of two particles.  

\subsection{Cuts and distributions} 
The \verb|Cut| item of Menu 1 allows the  user to apply cuts to the
current process.  This is done by filling in the cuts table with the
kinematical function (second column), the minimum limit for the cut
(third column) and the maximum limit for the cut (fourth column).  The
minimum and maximum limits for the cut should be algebraic functions
of the model parameters and numbers.  One of the limits can be left
blank, in which case that limit is not applied.  If there is an
exclamation mark (\verb|!|) in the first column, the cut is inverted.  
For example
\begin{verbatim}
  ! |Parameter   | Min bound   |Max bound
    | M(b,B)     | Mh/2        | Mh*2        
\end{verbatim}
means that the invariant mass of a b-quark and an anti-b-quark must be
between half the Higgs mass and twice the Higgs mass.

\begin{figure}[h]   
\includegraphics[width=0.9\linewidth]{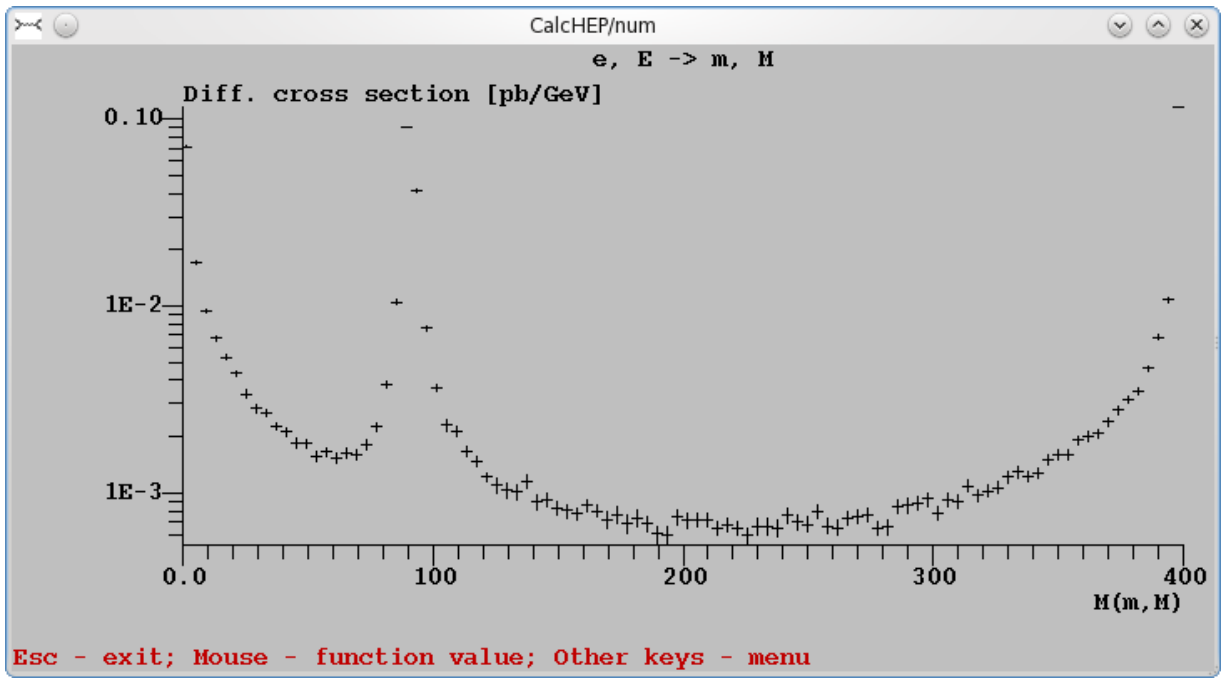}
\caption{\label{mm_plot} Plot of the invariant mass of a muon and an
  anti-muon for the process
e,E$\to$m,M  with an initial energy of  200 GeV and ISR   energy
smearing turned on.}
\end{figure}

The syntax for the distribution table [Menu 6] is similar to that of the cut table.
Both one- and two-dimensional distributions are supported. 
The distributions are filled during Monte Carlo integration and can be
viewed graphically with the \verb|Display distribution| item of Menu 6.
An example of a distribution in the GUI is presented in
Fig.~\ref{mm_plot}. 
These distributions can be saved in a few different formats.  The
first is to the file \verb|plot_n.txt| where \verb|n| enumerates the
plot files.  This file has the data in two space separated columns.
It also contains example instructions at the top for using this file with PAW and
Gnuplot.  
The data in these files can be  re-displayed later  by \CalcHEP~by the
\verb|plot_view| routine as in
\begin{verbatim}
   $CALCHEP/bin/plot_view plot_N.txt
\end{verbatim}
The \verb|Clear Statics| function of Menu 6 resets all the data. 

We note that the kinematic functions used in the \verb|Cut| and \verb|Distribution| tables can
contain particles  which  are not present in the current
subprocess. When this occurs, these cuts are ignored and these
distributions are not filled.
In this way, cuts and distributions can be defined once for all
subprocesses.  When a different subprocess is chosen [Menu 1], the
cuts and distributions that apply are automatically turned on.



\subsection{Monte-Carlo simulation and event generation}
\label{MC}

\CalcHEP~uses VEGAS \cite{Lepage, NumRec} for Monte Carlo integration. 
The \verb|Start Integration| function of Menu 6 launches  a cycle of \verb|NSess| VEGAS
sessions with \verb|nCalls| integrand calls for each session. If the switch
\verb|Freeze grid = OFF|,  VEGAS improves the integration grid
after each session\footnote{The importance sampling algorithms is used here.}.
Table \ref{tab:VegasSession}(left) presents an example of a Vegas
session with the grid being improved as evidenced by the steep decline
in the Monte Carlo uncertainty in the \verb|Error| column.
Although the grid is improved by the end of the session, the final
uncertainty (in the final line) is still large because it includes the
results of all ten sessions, including the initial sessions when the
grid was not yet improved.  It is a good idea to clear the
results once the grid is improved and rerun VEGAS.  This is done by using the
\verb|Clear statistics| menu item and then \verb|Start Integration|
again.  An example of a VEGAS session after the grid has been improved
and the statistics cleared is shown in Table
\ref{tab:VegasSession}(right).  The Monte Carlo uncertainty is small
for all ten sessions and the resulting cross section has a
correspondingly small uncertainty.

\begin{table}[h]
{\footnotesize
\begin{tabular} {|c|c|c|c|c||c|c|c|c|c|c|}
\hline
  IT & Cross section[pb]& Error[\%] & nCalls & $\chi^2$   &       IT& Cross section[pb] &Error[\%]   &  nCalls  &   Eff.   & $\chi^2$\\
\hline
   1 &    1.8319E+00    & 1.29E+01  &   9826 &       &            1 &    2.0516E+00 &  2.33E-01 &    9826  &          &  \\
   2 &    2.0575E+00    & 5.76E+00  &   9826 &       &            2 &    2.0587E+00 &  2.27E-01 &    9826  & 8.5E-01  &\\
   3 &    1.9409E+00    & 4.88E+00  &   9826 &       &            3 &    2.0506E+00 &  2.27E-01 &    9826  & 7.5E-01  &\\
   4 &    2.2081E+00    & 6.63E+00  &   9826 &       &            4 &    2.0450E+00 &  2.28E-01 &    9826  & 7.1E-01  &\\
   5 &    2.0717E+00    & 2.44E+00  &   9826 &       &            5 &    2.0511E+00 &  2.27E-01 &    9826  & 6.9E-01  &\\
   6 &    2.0588E+00    &  7.72E-01 &    9826&       &             6&     2.0549E+00&   2.26E-01&     9826 &  6.7E-01 &\\
   7 &    2.0624E+00    &  3.93E-01 &    9826&       &             7&     2.0623E+00&   2.29E-01&     9826 &  6.6E-01 &\\
   8 &    2.0501E+00    &  2.81E-01 &    9826&       &             8&     2.0610E+00&   2.28E-01&     9826 &  6.5E-01 &\\
   9 &    2.0472E+00    &  2.48E-01 &    9826&       &             9&     2.0516E+00&   2.31E-01&     9826 &  6.5E-01 &\\
  10 &    2.0547E+00    &  2.24E-01 &    9826&       &            10&     2.0564E+00&   2.27E-01&     9826 &  6.4E-01 &\\
 \verb|<>|    &     2.0383E+00   & 1.58E+00  &  98260 &     0.9 & \verb|<>|  &     2.0543E+00&   7.23E-02&    98260 &  6.4E-01 &   1\\
\hline
\end{tabular}
\label{tab:VegasSession}
\caption{ Result of a cycle of 10 Vegas sessions 
for the process e,E$\to$m,M with an total energy of 400 GeV beams and
ISR smearing turned on.
The left side corresponds with an intial cycle where the grid is
improved and {\tt Freeze grid=OFF}.  The right side corresponds with
a second cycle after the grid has been improved and the statistics
have been cleared.  {\tt Freeze grid=ON} has also been set in the
second cycle.  As a result, the generator grid is also being improved.}
}
\end{table}

When \verb|Freeze grid = ON| [Menu 6], VEGAS will prepare the event
generator during the VEGAS sessions.  With each VEGAS pass, the event
generator will improve its estimates of the maximum differential cross
section (or differential partial width) in each event generation
cube.  The number of event generation cubes can be set with the
\verb|Event Cubes| item of [Menu 6] as in
\begin{verbatim}
     Freeze grid = ON 
     Event Cubes  10000
\end{verbatim}  
The larger the number of event
cubes, the longer it takes to improve the event generator but the more
efficient the generator.  In each cube, the maximum value will be set
at $1.2$ times the largest differential cross section (or differential
partial width) that VEGAS finds.  Since with each pass the estimates
for the maximums will become  larger (as VEGAS finds higher
values), the event generator efficiency will become smaller.  The
event generator is prepared when the efficiency stabilizes. An example
of event generator preparation is shown in Table
\ref{tab:VegasSession}(right), where the \verb|Eff| column gives the
estimated efficiency of the event generator after each VEGAS session.

The event generator preparation in the above paragraph is based on the
Von Neumann  algorithm (see \cite{ParticlesFields}, p.202) where
a phase space point $x$   is sampled  with a probability  $F(x)$
(where in our case $F(x)$ is the maximum deteremined during the event generator preparation) and  
accepted with probability $f(x)/F(x)$ (where for us $f(x)$ is the
differential cross section or partial width at $x$). 
However, because the maximums are determined by the calculation of
random phase space points (by the VEGAS Monte Carlo), there are times
when $f(x)>F(x)$ will occur during event generation.  
If \CalcHEP~finds such a point $x$, it will increase $F(x)$ for the
the rest of the event generation. 
Furthermore, if \verb|Regenerate events| [Menu 7] is set to
\verb|OFF|, \CalcHEP~will accept this event and give it the weight
$f(x)/F(x)>1$.  On the other hand, if \verb|Regenerate events| is set
to \verb|ON|, \CalcHEP~will regenerate all the events from that event
cube.
When \CalcHEP~is finished generating the requested events, it will
display information about the events on the screen as in
\begin{verbatim}
          Event generated: 10000
            efficiency:  6.4E-01
       Max Event multiplicity: 2
       Multiple events(total): 3
\end{verbatim}
including whether events with weight greater than 1 were generated
 and request whether the user would like to accept the
generated events.

\section{Working with files from numerical session}\label{collecting}   

\subsection{File naming convention}
\CalcHEP~keeps an internal counter of numerical sessions.  Any time a
parameter is changed that affects the numerical results,
\CalcHEP~increments the session number.  All the parameters as well as
the state of the random number generator and the integration results
are stored in the file \verb|$WORK/results/session.dat|.  This allows
the numerical session to be run later starting from the same state.
The session number is displayed on the screen in the GUI.
Furthermore, the session number is used in the names of some files
written by \CalcHEP~in order to connect them with a particular
session.  These files are
\begin{itemize}
\item  \verb|prt_N| which contains complete information about the
  session parameters (including model parameters, momenta, structure
  functions, cuts, etc.) as well as some of the results from the VEGAS
  session;
\item \verb|decaySLHA_N| which contains information about the particle
  spectrum of the model and includes the particles'
decay channels in the format of \cite{Skands:2003cj} 
\item \verb|distr_N| which contains the full distribution data  filled
  by  VEGAS which can be viewed later as described in Section
  \ref{sec:display distributions};
\item \verb|events_N.txt| which contains the generated events;
\end{itemize}
where \verb|N| is the session number.

\subsection{\label{sec:display distributions}
Distribution files}
The  distributions generated by \CalcHEP~can be displayed after a numerical
session has finished with the command
\begin{verbatim}
        $CALCHEP/bin/show_distr distr_N
\end{verbatim}
where \verb|N| is the session number where the distribution was generated.
The distributions generated in different numerical sessions can be
combined   by the command
\begin{verbatim}
       $CALCHEP/bin/sum_distr distr_A distr_B ... > distr_sum
\end{verbatim}
where \verb|distr_A, distr_B ...| are the distribution files from
different sessions and \verb|distr_sum| is the file where the results
should be written.  
This program only combines distributions with exactly the same
kinematical variable and exactly the same distribution limits.  For
example, 
\verb|M(b,B)| and \verb|M(B,b)| are treated   as different
distributions are never summed.
However, the distribution  \verb|M(jet,jet)|  where the alias name
\verb|jet| has been defined  appropriately (see Section \ref{sec: num alias})  are combined if their
distribution limits are identical as well.

\subsection{Events} 
\CalcHEP~writes events in the format presented in
\cite{Pukhov:1999gg}. 
However, the the  LHE format
\cite{Alwall:2006yp} has become widely used now.  For this reason, we
include a script to rewrite event files in LHE format which can be run
as in:
\begin{verbatim}
       $CACLHEP/bin/event2lhe events_N.txt  > events_N.lhe
\end{verbatim}
where \verb|N| is the session number for the generated events.
Additionally, \CalcHEP~has routines to read the event files and
histogram the events.  The notation for the kinematical observables is
the same as in Section \ref{functions}.  These routines can be run as
in the following examples:
\begin{verbatim}
       $CALCHEP/bin/events2tab Variable Min Max Nbin < events.txt > tab.txt
       $CALCHEP/bin/lhe2tab    Variable Min Max Nbin < events.lhe > tab.txt
\end{verbatim}
where \verb|Variable| is the kinematic observable and must be in
quotation marks (e.g. \verb|"M(m,M)"|), \verb|Min| and
\verb|Max| are the histogram's minimum and maximum values respectively, \verb|Nbin|
is the number of bins, \verb|events.txt| and \verb|events.lhe| are the
event files in original and LHE format respectively and \verb|tab.txt|
is the file where the results should be written.
In the case of the \verb|lhe2tab|, the PDG ID's of the particles
should be used in place of the particle's names since the particle
names are absent in the LHE format.
 The resulting histograms can be displayed on the screen and transformed into
PAW, Gnuplot, Mathematica and \LaTeX~formats  by the  \verb|plot_view| routine.   
For further analysis  \CalcHEP~contains a program that creates PAW
NTUPLES from LHE files which is used in the following way
\begin{verbatim}
       $CALCHEP/bin/nt_maker events.lhe
\end{verbatim}
where \verb|events.lhe| is an event file in LHE format.

\subsection{Event Mixing and LHE format}\label{lhef}

Typical collider processes contains many subprocesses that differ only by the initial state and/or final state
particles.  For example, at the LHC, the initial states are two
colliding protons which, however, are composed of quarks, antiquarks
and gluons. It is  desirable to combine different channels in one event  file and  connect
production events with decays so that the final events are fully or
partial decayed.  The \CalcHEP~routine  which does this job is \verb|event_mixer| and can be used as in
\begin{verbatim}
        $CALCHEP/bin/event_mixer Lumi Nevent  dir1  dir2  ... 
\end{verbatim}
where \verb|Lumi| is the maximum integrated luminosity (in units of $1/fb^{-1}$), \verb|Nevents| is the number of events to generate, and \verb|dir1 dir2 ...| are the directories where the production and decay events are stored.  If \verb|Nevents| is smaller than \verb|Lumi| times the final cross section (we will call this $N_{max}$), then \verb|Nevents| will be produced.  If, on the other hand, \verb|Nevents| is larger, then \verb|event_mixer| will stop when it reaches $N_{max}$.
If \verb|event_mixer| comes to the end of a decay file before it is finished producing the requested number of events, it prints a message to stderr and returns to the beginning of the decay event file.
The resulting events are stored in th file \verb|event_mixer.lhe| in LHE format
\cite{Alwall:2006yp}.
      
Before \verb|event_mixer| begins combining events, it reads the  file \verb|decaySLHA.txt| which should be stored in the current
directory. This file should  contain the quantum numbers, masses, widths and branching ratios of the particles of the model
  written in SLHA format \cite{Skands:2003cj}. This file is used to define the
correct widths and branching ratios of the decaying particles.
If this file is not present, \verb|event_mixer| will determine the
branching ratios from the decay event files that it uses.  However,
please note that if \verb|decaySLHA.txt| is not present and all the
decay channels for an unstable particle are not present,
\verb|event_mixer| {will likely produce incorrect results.
The
\verb|decaySLHA.txt| file can be produced  by the \CalcHEP~numerical
session (see Section \ref{sec:decaySLHA}).  We strongly recommend to always include it when using \verb|event_mixer|.}

After reading \verb|decaySLHA.txt| but before combining events, \verb|event_mixer| prints to stdout the final cross
section and the maximum number of events that can be generated.  For example,
\begin{verbatim}
2.368E-01  -total cross section[pb]
10098      -maximum number of events
\end{verbatim}
To get this information before mixing the events, simply request $0$ events.

Some special features of LHE file generated by \CalcHEP~are:
\begin{itemize}

\item The decays are applied recursively. 

\item  A  decay history of each event is stored in the LHE file.  
This includes information about  
       the parent particles and their mean life time.  
This information
can be used for proper hadronisation and  detector simulation.

\item When connecting decays, \verb|event_mixer| uses a Breit-Wigner virtual mass
distribution, where we assume that the matrix elements of the subprocesses do not depend
strongly on the off-shell momentum. 

\item If \verb|decaySLHA.txt| file was detected it is  attaches  to the output 
 inside  \verb|<slha> ... </slha>| tags according to the LHE convention. 
This allows parton shower generators like
Pythia and Herwig to implements the decays of BSM particles.

\item In case of an incomplete set of decay channel event files (which is detected  via a
difference between the sum of  the partial widths from the decay event files and that stored in \verb|decaySLHA.txt|)
the  resulting cross section is reduced correspondingly. 

\item  In spite of   Breit-Wigner mass smearing for decays  our procedure does not break momentum
conservation.  The output file  contains a line which records the largest
deviation from energy momentum conservation.  An example is
\begin{verbatim}
   #lost_momenta_max/Etot 7.9E-11 1.3E-12 1.3E-12 8.0E-11 
\end{verbatim}
Typical value should be on the order of $10^{-10}$ since the original event files contain
 11 digits Of precision for the particle  momenta. 

\end{itemize}

The generated \verb|event_mixer.lhe| file also contains
an  XML header spanned by the tags  \verb|<hepml>| \verb|..</hepml>| written  in {\tt HepML} \cite{Belov:2010xm}
format. This  allows to automatically upload the LHE file to the 
CERN Monte-Carlo Database (MCDB) using the command\footnote{ The {\tt upload2mcdb\_hepml.pl} script can be downloaded from
the MCDB website {\tt https:\textbackslash\textbackslash mcdb.cern.ch.}}
\begin{verbatim}
    ./upload2mcdb_hepml.pl -header hepml event_mixer.lhe   
\end{verbatim}
If the file \verb|run_details.txt| is found in the current directory,
\verb|event_mixer| will include the information in this file in the
header.  The format for this file is a keyword value pair on each
line and is the same as the format for a batch file used with the
batch interface.  Further details can be found in Section~\ref{Batch}.

\section{CalcHEP blind mode and batch scripts} \label{Batch}

Initially \CalcHEP~was designed for interactive calculations with a
graphical user interface.  However, there are times when a batch
system is ideal.  For example, when a calculation takes a very long
time, or the user is interested in doing scans over parameter space
or over subprocesses.

In order to solve this problem, the \verb|blind| mode was introduced
\cite{Pukhov:2003,Pukhov:2004ca}.  If the \verb|-blind| flag is used
with either \verb|s_calchep| or \verb|n_calchep|, they will read the
next argument and interpret it as a series of commands.  These
commands are written in a special notation which matches the
keystrokes the user would perform during an interactive session.  For
example, the command\footnote{One has to remove tmp/safe file before
launching this command.}
\begin{verbatim}
    $CALCHEP/bin/s_calchep -blind "fStandard Model{{{e,E->m,M{{[{[{{0"
\end{verbatim}
would generate the C code for the process \verb|e,E->m,M|
in the \verb|Standard Model|.  
Some of the characters in the command string have a special meaning.
Here are a few of them:
\begin{center}
\begin{tabular}{|cl|cl|}
\hline
\verb|]| & -Up      &  \verb|}| & -Escape  \\
\verb|[| & -Down    &  \verb|f| & -find or search for the string in a menu\\
\verb|{| & -Enter   &  \verb|0-9|  & -Function keys or numeric input  according to context\\
\hline
\end{tabular}
\end{center}

Determining an appropriate command sequence string can be difficult
for a user.  For this reason, both \verb|s_calchep| and
\verb|n_calchep| can be run with the flag \verb|+blind|.  In
\verb|+blind| mode, the interactive graphical interface opens and the user can run
them as usual.  When the user quits, \CalcHEP~writes the command
sequence string to stdout.  The user can then copy it and modify as
appropriate to use in \verb|-blind| mode.  
Further technical   details
can be  found  in \cite{Pukhov:2003,Pukhov:2004ca}. 

The \verb|-blind| mode was used to write
several scripts which are stored in the \verb|$CALCHEP/bin| 
directory. First of all, there are several scripts which change the
parameters of a
numerical session:\\
{$\bullet$\verb|set_momenta p1 p2|:}  This script updates the
  momenta of the incoming particles to \verb|p1| and \verb|p2| and
  then quits.\\
{$\bullet$\verb|set_param name1 value1 name2 value2 ...|:} This
  script changes the numerical values of one or more of the independent model
  parameters \verb|name1|, \verb|name2|, etc. to \verb|value1|,
  \verb|value2|, etc. respectively and then quits.\\
{$\bullet$\verb|set_param File|:} In this case, this script
  changes the numerical values of the independent model parameters
  as specified in the file \verb|File|.  \verb|File| must have each
  model parameter on a separate line with the name coming first
  followed by the new numerical value, separated by white space.\\
{$\bullet$\verb|set_vegas nSess1 nCalls1 nSess2 nCalls2 EventCubes |:}
   This script sets the  parameters of a two loop Vegas  calculation
   used by the 
   \verb|run_vegas| script presented below.  The meaning of the
   parameters was explained in Section \ref{MC}.  They can be seen in 
   Fig.\ref{num_module}.  The statistics are cleared and the grid is frozen between session
   1 (defined by \verb|nSess1| and \verb|nCalls1|) and session 2
   (defined by \verb|nSess2| and \verb|nCalls2|).

The full set of parameters for the numerical session are stored 
in the file \verb|session.dat|.  When the interactive GUI session
(\verb|n_calchep|) runs or when any of these scripts run, they update
\verb|session.dat| with the new parameters before quitting.  As a
result, when the interactive GUI session (\verb|n_calchep|) or these
scripts are run later, they begin with the updated parameters of the
last session.  This allows the user to prepare for a blind calculation
in two ways.  The user can either run the interactive GUI session
(\verb|n_calchep|) and set all the parameters as needed or \heshe~can
run these scripts to set the parameters as required.  Once the
parameters are set appropriately, the folowing scripts allow to run
VEGAS:\\
{$\bullet$ \verb|run_vegas|:}
This script runs VEGAS according to the parameters in \verb|session.dat|.

There are  several scripts which perform scans which are based on
\verb|run_vegas|.  The output  for these cycles is  stored in text files whose names
have the form \verb|xxx_j1_j2|, where \verb|j1| is
the session number when the script began and \verb|j2| is the session number when it
finished. If there are distributions specified in \verb|session.dat|,
  they are stored in the files \verb|distr_k| where  \verb|j1|$\le$ k$\le$
\verb|j2|. These scripts are:\\
{$\bullet$\verb|pcm_cycle pcm0 step N|:}  This
  script scans the cross-section over the center of mass energy.  For each point in the
  scan, it updates the momenta of the initial state particles and then runs the Vegas Monte
  Carlo integration. It begins its calculations with the  momenta  of the initial state particles
  equal to \verb|pcm0| and increases in steps of size \verb|step| for a total of \verb|N| steps. 
  When it is finished, it writes the resulting cross-sections to the file
  \verb|pcm_tab_j1_j2|.\\
{$\bullet$\verb|name_cycle name val0 step N|:} This script scans
  the cross-section over a model parameter's value.  For each point in
  the scan, it updates the parameter \verb|name| and then calculates
  the cross-section.  When it is finished, it writes the resulting
  cross-sections to the file \verb|name_tab_j1_j2|.
  where \verb|name| is the name of the parameter.\\
{$\bullet$\verb|subproc_cycle L Nmax|:} This script calculates
  the cross-section and generates events for each subprocess.  When it
  is finished, it adds the cross-sections together and prints the
  total cross-section to the screen.  If there are distributions
  specified then they are added together and the resulting distribution is
  stored in the file \verb|distr_j1_j2|.  It
  also generates unweighted events for each subprocess.  The number
  it generates is equal to the smaller of \verb|Nmax| and the cross-section times the luminosity
  which is specified by \verb|L|.  It writes these
  events to the files \verb|events_k.txt| where \verb|j1|$\le$ k$\le$ \verb|j2|.\\
{$\bullet$\verb|par_scan < data.txt|:}  This script calculates
  the cross-sections according to the grid for names and parameters
  given in \verb|data.txt| file. The format of \verb|data.txt|
  is
{\footnotesize
  \begin{verbatim}
  name_1  name_2 ... name_N
  val_11  val_12 ... val_1N
  .........................
  val_M1  val_M2 ... val_MN
  \end{verbatim}
}
\noindent
  where \verb|name_1  name_2 ... name_N| are the names of independent model parameters
 , while \verb|val_11| \verb|... val_1N| are the values for the respective
  parameters to be  used for the first calculation and 
   \verb|val_M1 ... val_MN| are the values for these parameters for
   the last parameter point to be calculated.
  Note that this script does not sum over the subprocesses 
  (i.e. it will do the calculation only for the subprocess currently
  chosen in \verb|session.dat|).
   The results of the calculations are printed to stdout and  can be
   redirected into a file with a command such as
\begin{verbatim}
   par_scan < data.txt > results.txt
\end{verbatim}
where \verb|results.txt| is the file where the results should go.
   The output format repeats the input format but contains one
   additional column with the results of the calculation.\\
{$\bullet$\verb|par_scan_sum < data.txt|:} This script calculates
the cross-sections according to the parameter points in
\verb|data.txt| similarly to \verb|par_scan|.  However, this script
calculates the cross section for all the available subprocesses and
sums them together.\\
{$\bullet$\verb|gen_events Nevents|:}  This script can by launched
after a successful VEGAS calculation including a session where the
grid is frozen to improve the event generation grid.  This can be done
with the \verb|run_vegas| script described in this section.  The
argument \verb|Nevents| defines the number of events to generate.

If any of these scripts ends with an error, a message is printed to
\verb|stderr| and the return value of the script can be
seen by issuing \verb|echo $?| on the shell.   A description of the
possible error codes can be found in the \CalcHEP~manual.


\section{\label{sec:batch interface}Batch interface}

Although the shell scripts of the previous subsection greatly improve
the users ability to run their desired processes in batch mode, there
are still some limitations when doing large complex calculations
involving scans over parameter space, many subprocesses and
parallelization.  To overcome these challenges, we have written a Perl
script which we call the ``batch interface''.  The main
features of this Perl interface are:
\begin{itemize}
\item The input is a pure text file we call the ``batch file''.  It
  consists of a series of keywords together with values for those
  keywords, with each keyword on a separate line.  Most of the options
  available in the interactive session are supported by keywords in
  the batch file and thus most calculations can be done using the
  batch interface.
\item A library of subprocess numerical codes is utilized.  Each time
  the batch interface is run, it first checks whether the subprocess
  numerical code exists.  If it does, it reuses it and skips the
  often long process of code generation.  Any requested numerical
  codes not in the library are then generated and added to the library.  If
  the model changed, the numerical codes are regenerated as
  appropriate.
\item The numerical phase space integration is done and events are
  generated for each subprocess and the results are combined.
  Production and decay events are connected and the final event output
  is an LHE file with all the events fully decayed which can be used
  directly by Pythia or other software.
\item Multiple parameters can be scanned over.  For each parameter
  point, the results are combined and stored with names unique to that
  parameter point for easy retrieval.
\item Both the symbolic calculations and the numerical calculations
  are parallelized.  Each subprocess and each parameter point are run
  as separate jobs and run on all available cpu cores.  The number of
  cores available is set by the user as is the type of cluster
  software used.  Multicore machines, PBS cluters and LSF clusters are
  currently supported.
\item The progress of the calculation is stored in a series of html
  files which can be viewed in a web browser\footnote{The inspiration
    for creating html pages that showed the progress of the
    calculation was obtained from MadGraph \cite{Alwall:2007st}.}.  These html pages
  contain information about the progress of the calculation as well as
  the results of the calculations which are already finished.  The
  final event files are linked as are the session.dat and prt files
  which give the full details of each individual calculation.  Pure
  text versions of the progress pages are also created for situations
  where a web browser is not convenient.
\end{itemize}
Once the user creates the batch file and runs the batch interface, no
user input is required until it finishes.  It can be run in the
background and checked periodically.

After the user has created their batch file, they would typically run
the batch interface from their \CalcHEP~work directory as
\begin{center}
\verb|./calchep_batch batch_file|
\end{center}
where \verb|batch_file| is the name of their batch file, which can be
named anything the user likes.  The batch interface will start by
printing a message to the shell which will contain the location of the
html progress reports which the user can simply copy and paste into
their browser url window.  The first time the user runs the batch
interface, they can also run the following from the work directory
\begin{center}
\verb|./calchep_batch -help|
\end{center}
which will complain that no batch file was present, create a series of
html help files and quit.  The location of the html help files will be
printed to screen.  This html help file can be opened in a web browser
and contains all the details that are presented here.  

{In the following subsection} we  describe each keyword available
for the batch file and how to use it. An example batch file is stored
in \verb|CALCHEP/utile/batch_file|.

\subsection{Batch files}

\subsection*{Comments}
Any line beginning with a \verb|#| is ignored by run\_batch. The \verb|#| has to be
at the very beginning of the line. Some examples are:
\begin{verbatim}
# This is ignored.
#Model: Standard Model	 This is ignored.
Model: # Standard Model(CKM=1) This is not ignored.
\end{verbatim}

\subsection*{Model}
The first section of the batch file should contain the specification
of the model. This is done by model name and should match exactly the
name in the \CalcHEP~model list. So, if you want to run the ``Standard
Model(CKM=1)", you would specify this with the batch file line:
\begin{verbatim}
Model	:	Standard Model(CKM=1)
\end{verbatim}
There is no default for this line. It must be included.

The gauge of the calculation should also be specified in this section. Choices are Feynman and unitary gauge. \CalcHEP~is much better suited to calculation in Feynman gauge, but there may be times that unitary gauge is useful. This can be specified using the keyword \verb|Gauge| as in:
\begin{verbatim}
Gauge	:	unitary
\end{verbatim}
The default is \verb|Feynman|.

\CalcHEP~allows decays of particle such as the Higgs boson via
off-shell W and Z bosons (see Section~\ref{symbolic_module}).  This behavior can be controled by the key phrase \verb|Virtual W/Z decays| as in:
\begin{verbatim}               
Virtual W/Z decays : Off
\end{verbatim}
The default is \verb|On|.

\subsection*{Process}
Processes are specified using the \verb|Process| keyword and standard
\CalcHEP~notation as in:
\begin{verbatim}
Process	:	p,p->j,l,l
\end{verbatim}
Multiple processes can also be specified as in:
\begin{verbatim}
Process	:	p,p->E,ne
Process	:	p,p->M,nm
\end{verbatim}
As many processes as desired can be specified. When more than one process is specified, the processes are numbered by the order in which they are specified in the batch file. So, in this example, \verb|p,p->E,ne| is process 1 and \verb|p,p->M,nm| is process 2. This numbering can be useful when specifying QCD scale, cuts, kinematics, regularization and distributions allowing these to be specified separately for each process. There is no default for this keyword. It must be specified.

Decays are specified using the \verb|Decay| keyword and are also in
standard \CalcHEP~notation as in:
\begin{verbatim}
Decay	:	W->l,nu
\end{verbatim}
Again, multiple decays can be specified as in:
\begin{verbatim}
Decay	:	W->l,nu
Decay	:	Z->l,l
\end{verbatim}
The default is to not have any decays. Cuts, kinematics, regularization and distributions do not apply to decays.

It is sometimes convenient to specify groups of particles as in the
particles that compose the proton or all the leptons. This can be done
with the keyword \verb|Alias| as in:
\begin{verbatim}
Alias	:	p=u,d,U,D,G
Alias	:	l=e,E,m,M
Alias	:	nu=ne,Ne,nm,Nm
Alias	:	W=W+,W-
\end{verbatim}
As many aliases as necessary can be specified. These definitions can
be used in cuts and distributions as well as in the processes and
decays. The default is not to have any alias definitions.

\subsection*{PDF}
The PDF of a proton or antiproton can be specified with the
\verb|pdf1| and \verb|pdf2| keywords which correspond to the pdfs of
the first and second incoming particles respectively. Choices for
these keywords are:
{\footnotesize
\begin{verbatim}
   cteq6l (proton)         cteq6l (anti-proton)       
   cteq6m (proton)         cteq6m (anti-proton)       
   cteq5m (proton)         cteq5m (anti-proton)       
   mrst2002lo (proton)     mrst2002lo (anti-proton)   
   mrst2002nlo (proton)    mrst2002nlo (anti-proton)  
   None
\end{verbatim}
}
An example for the LHC is:
{\footnotesize
\begin{verbatim}
pdf1	:	cteq6l (proton)
pdf2	:	cteq6l (proton)
\end{verbatim}
}
The default is \verb|None|. These keywords can also be used for electron positron colliders. For this process the available pdfs are:
{\footnotesize
\begin{verbatim}
  ISR
  ISR & Beamstrahlung
  Equiv. Photon
  Laser photons
  None 
\end{verbatim}
}
The following proton electron collider pdf is also available:
{\footnotesize
\begin{itemize}
\item \verb|Proton Photon|
\end{itemize}
}
All of these pdfs must be typed exactly or copied into the batch file.

If \verb|ISR & Beam| is chosen, then the following beam parameters may
be specified:
{\footnotesize
\begin{verbatim}
Bunch x+y sizes (nm)	:	550
Bunch length (mm)	:	0.45
Number of particles	:	2.1E+10
\end{verbatim}
}
The default values are the default values in \CalcHEP~and correspond roughly with the ILC.

If \verb"Equiv. Photon" is chosen for the pdf, then the following
parameters may be specified:
{\footnotesize
\begin{verbatim}
Photon particle	:	e^-
|Q|max	:	150
\end{verbatim}
}
Choices for the \verb"Photon particle" keyphrase are \verb|mu^-|,
\verb|e^-|, \verb|e^+|, \verb|mu^+|. The default is \verb|e^+|. The
default for the keyword \verb"|Q|max" is the same as in the CalcHEP
interactive session.

If \verb"Proton Photon" is chosen then the following may be specified:
{\footnotesize
\begin{verbatim}
Incoming particle mass	:	0.937
Incoming particle charge	:	-1
|Q^2|max	:	2.1
Pt cut of outgoing proton	:	0.11
\end{verbatim}
}
The defaults are the same as in the \CalcHEP~interactive session.

The user can also use PDFs from the SLHA library if it is installed
and  the respective link is given in the model \verb|Libraries| table.
Here is an example of how to use PDF functions from the LHAPDF library:
{\footnotesize
\begin{verbatim}
pdf1: LHA:cteq6ll.LHpdf:0:1     
pdf2: LHA:cteq6ll.LHpdf:0:1      
\end{verbatim}
}
In this case, the value for \verb|pdf1| and \verb|pdf2| should be
constructed using the following rules:
\\
1) It should begin with \verb|LHA|.
\\
2) \verb|LHA| should be followed by the name of a particular PDF set
located in the \verb|PDFsets| directory (e.g. \verb|cteq6ll.LHpdf| in the example above).
\\
3) The PDF set name should be follwoed by  a particular PDF set number
(e.g. \verb|0| in the example above corresponding with the central fit).
\\
4) This should be followed by either  \verb|1| (for a proton) or  \verb|-1|
(for an antiproton).
\\
5) The pieces from instructions  (1)-(4) should be separated by a colon (\verb|:|).

\subsection*{Momenta}
The momenta of the incoming states can be specified with the keywords
\verb"p1" and \verb"p2" and are in GeV as in:
{\footnotesize
\begin{verbatim}
p1	:	7000
p2	:	7000
\end{verbatim}
}
These are the default values for the momenta.

\subsection*{Parameters}
The default parameters of the model are taken from the varsN.mdl file
in the models directory. Other parameter values can be used if
specified using the \verb"Parameter" keyword. Here is an example:
\begin{verbatim}
Parameter	:	EE=0.31
\end{verbatim}
This gives a convenient way of changing the default values of the parameters. Simply open \CalcHEP~in symbolic mode, choose to edit the model and change the values of the indepenedent parameters. These new values will then become the default values used by this batch program. There is no need to redo the process library.

\subsection*{Scans}
In some models it is useful to scan over a parameter such as the mass
of one of the new particles. For example, if there is a new W' gauge
boson, it may be desireable to generate events and/or distributions
for a range of masses for the W'. This can be done with the 
\verb"Scan parameter", \verb"Scan begin", \verb"Scan step size" and 
\verb"Scan n steps" keyphrases. Here is an example:
{\footnotesize
\begin{verbatim}
Scan parameter	:	MWP
Scan begin	:	400
Scan step size	:	50
Scan n steps	:	17
\end{verbatim}
}
This will generate the events and/or distributions for the model with
the mass of the W' set to 400GeV, 450GeV, 500GeV,...1200GeV. As many
scans as desired can be specified (including zero). For each scan, all
four keyphrases have to be specified. Furthermore, if there is more
than one scan, all four keyphrases have to be specified together.

\subsection*{QCD}
The parameters of the QCD menu of the numerical session can be
specified as in the following example:
{\footnotesize
\begin{verbatim}
parton dist. alpha	:	ON
alpha(MZ)	:	0.118
alpha nf	:	5
alpha order	:	NLO
mb(mb)	:	4
Mtop(pole)	:	174
alpha Q	:	M45
\end{verbatim}
}
The default values are the ones in the interactive session. Not all
the keywords have to be included in the batch file. It is sufficient
to include the ones that need to be changed. 

The QCD scale can be specified in terms of the invariant mass of
certain final state particles as in \verb"Mij" which means that the
QCD scale is taken to be the invariant mass of particles \verb|i| and
\verb|j|. Or, it can be specified as a formula in terms of the
parameters of the model as in \verb"Mt/2" which means half of the top
quark mass. When specifying the scale in terms of the invariant mass
of final state particles, the numbers are taken from the way the
processes are entered with the \verb"Process" keyword. So, if the
process is specified as \verb"p,p->j,l,n", \verb"M45" means the
invariant mass of the lepton and neutrino (l,n). The batch script will
take care of renumbering if the subprocesses have the final state
particles in a different order. It is also sometimes useful to use a
different scale for different processes. For example, suppose the two
processes \verb"p,p->j,l,n" and \verb"p,p->j,j,l,n" are specified in
the batch file, the scales could be specified as in this example:
{\footnotesize
\begin{verbatim}
alpha Q	:1:	M45
alpha Q	:2:	M56
\end{verbatim}
}
The number between the \verb|::| specifies which process to apply this
scale and corresponds to the order in which the user specified the
processes. If more than one process is specified, but the same non
default scale is desired for all of them, this can be specified as in:
{\footnotesize
\begin{verbatim}
alpha Q	:	Mt/2
\end{verbatim}
}
This specification will apply the same scale \verb"Mt/2" to all processes.

\subsection*{Cuts}
Cuts are specified with the keywords \verb"Cut parameter", 
\verb"Cut invert", \verb"Cut min" and \verb"Cut max" and use standard CalcHEP
notation, except for \verb"Cut invert" which can be either \verb"True"
or \verb"False". These cuts are only applied to the production
processes. They are not applied to the products of the decays. Here is
an example:
{\footnotesize
\begin{verbatim}
Cut parameter	:	T(le)
Cut invert	:	False
Cut min	        :	20
Cut max	        :	
\end{verbatim}
}
For each cut, all four keyphrases have to be present. As many cuts as
desired can be included. Including \verb"Cut min" or \verb"Cut max"
but leaving the value blank will leave the value blank in the CalcHEP
table. If the cut should only be applied to a certain process, then
the colon can be changed to \verb|:n:| where \verb|n| is the process
number.

\subsection*{Kinematics}
As the number of final state particles increases, it can be very
helpful to specify the \verb"kinematics" which helps \CalcHEP~in the
numerical integration stage. This is done in exactly the same notation
as in CalcHEP. The numbering corresponds to the order the particles
are entered in the process in the batch file. Here is an example:
{\footnotesize
\begin{verbatim}
Kinematics	:	12 -> 34 , 56
Kinematics	:	34 -> 3 , 4
Kinematics	:	56 -> 5 , 6
\end{verbatim}
}
If multiple processes are specified, using a single colon  will apply the kinematics to all processes. If
different kinematics are desired for each process, then the \verb|:n:|
notation can be used.

\subsection*{Regularization}
When a narrow resonance is present in the signal, it is a good idea to
specify the \verb"Regularization". This is done with the same notation
as in CalcHEP. Here is an example:
{\footnotesize
\begin{verbatim}
Regularization momentum	:	34
Regularization mass	:	MW
Regularization width	:	wW
Regularization power	:	2
\end{verbatim}
}
Regularization for as many resonances can be specified as
desired. Furthermore, different resonances can be specified for each
process using the \verb|:n:| notation.

\subsection*{Distributions}
Distributions are only applied to the production process. The decays
are ignored. Standard \CalcHEP~notation is used for the distribution
parameter. Here is an example:
{\footnotesize
\begin{verbatim}
Dist parameter	:	M(e,E)
Dist min	:	0
Dist max	:	200
Dist n bins	:	100
Dist title	:	p,p->l,l
Dist x-title	:	M(l,l) (GeV)
\end{verbatim}
}
The value for the keyphrase \verb"Dist n bins" has to be one of
{\footnotesize
\verb|300|, \verb|150|,	\verb|100|,	\verb|75|,	\verb|60|,	\verb|50|,
\verb|30|,	\verb|25|,	\verb|20|,	\verb|15|,	\verb|12|,	\verb|10|,
\verb|6|,	\verb|5|,	\verb|4|,	\verb|3| or	\verb|2|.
}
These are the values allowed by the \CalcHEP~histogram routines. The values given for the titles have to be pure text. No special characters are currently allowed. Gnuplot must be installed for plots to be produced on the fly and included in the html progress reports. More than one distribution can be specified. Also, distributions will work even if no events are requested.

For this to work, the distributions have to be unambiguous and apply to all subprocesses the same way. For example, if a process is \verb|p,p->l,l,l| and the distribution \verb|M(l,l)| is given, then this routine will not know which two leptons to apply the distribution to and the results are unpredictable. If the process is \verb|p,p->l,l| where \verb|l=e,E,m,M| and the distribution \verb|M(e,E)| is desired, this distribution will only apply to some of the subprocesses and give unpredictable results. Make sure your distribution is unambiguous and applies in exactly one way to each subprocess. If this is done, it should work. Nevertheless, check each distribution carefully to make sure it is being done correctly.

\subsection*{Events}
The number of events is specified with the keyphrase 
\verb"Number of events". This specifies the number of events to
produce after all subprocesses are combined and decayed. If a scan over
a parameter is specified, this keyphrase determines the number of
events to produce for each value of the scan parameter. The number of
events requested can be zero. In this case, the cross sections are
determined and the distributions generated but no events are
produced. Here is an example:
\begin{verbatim}
Number of events	:	1000
\end{verbatim}

The name of the file can be specified using the 
\verb"Filename" keyword. If specified, all the files will begin with
this name. Here is an example:
\begin{verbatim}
Filename	:	pp-ll
\end{verbatim}
 
If \verb|nt_maker| has been installed in the bin directory, PAW
ntuples can be made on the fly by setting \verb"NTuple" to \verb|True|
as in:
\begin{verbatim}
NTuple	:	True
\end{verbatim}
The default is \verb|False|.

The keyword \verb|Cleanup| determines whether the intermediate files
of the calculation are removed.  This can be useful if many large
intermediate files are created and space is an issue.  On the other
hand, it can be useful to keep the files when debugging is necessary.
If this keyword is set to \verb|True|, the intermediate files are
removed.  If set to \verb|False| then they are not removed.  Here is
an example:
\begin{verbatim}
Cleanup	:	False
\end{verbatim}

\subsection*{Parallelization}
The parallelization mode is set using the keyphrase 
\verb"Parallelization method" and can be either 
\verb"local", \verb"pbs" or \verb"lsf". In \verb|local| mode, the jobs
run on the local computer, in \verb|pbs| mode, the jobs are run on a
pbs cluster and in \verb|lsf| mode, the jobs are run on an lsf
cluster. If run from a pbs or lsf cluster, the terminal should be on
the computer with the pbs or lsf queue. Here is an example of setting
the batch to run in pbs mode:
\begin{verbatim}
Parallelization mode	:	pbs
\end{verbatim}
Local mode is the default.

If run in \verb|pbs| mode, there are several options that may be
necessary for the pbs cluster. All of them can be left blank in which
case they will not be given to the pbs cluster. Here is an example of
the options available:
{\footnotesize
\begin{verbatim}
Que	:	brody
Walltime	:	1.5
Memory	:	1
email	:	name@address
\end{verbatim}
}
The \verb|que| keyword specifies which pbs queue to submit the jobs to. \verb|Walltime| specifies the maximum time (in hours) the job can run for. If this time is exceeded, the jobs are killed by the pbs cluster. \verb|Memory| specifies the maximum amount of memory (in G) that the jobs can use. If this memory is exceeded by a job, the pbs cluster will kill the job. \verb|email| specifies which email to send a message to if the job terminates prematurely. The default for all of these is whatever is the default on the pbs cluster.

If run in \verb|lsf| mode, there is one more option in addtion to the those above:
\begin{verbatim}
Project	:	project_name
\end{verbatim}

\verb|Sleep time| specifies the amount of time (in seconds) the batch
script waits before checking which jobs are done and updating the html
progress reports. If a very short test run is being done, then this
should be low (say a few seconds). However, if the job is very large
and will take several hours or days, this should be set very high (say
minutes or tens of minutes or hours). This will reduce the amount of
cpu time the batch program uses. Here is an example setting the sleep
time to 1 minute:
\begin{verbatim}
sleep time	:	60
\end{verbatim}
The default is 3 seconds.

When jobs are run on the local computer, the keyword \verb|Nice level|
specifies what nice level the jobs should be run at. If other users
are using the same computer, this allows the job to be put into the
background and run at lower priority so as not to disturb the other
users. This should be between 0 and 19 where 19 is the lowest priority
and the nicest. Typically, it should be run at level 19 unless the
user is sure it will not disturb anyone. The nice level should be set
both for a local computer and for a pbs or lsf batch run. The reason
is that some jobs are run on the pbs or lsf queue computer even on the
pbs or lsf cluster. Here is an example:
\begin{verbatim}
Nice level	:	19
\end{verbatim}
Level 19 is the default.

\subsection*{Vegas}
The number of vegas calls can be controlled with the keywords 
\verb"nSess_1", \verb"nCalls_1", \verb"nSess_2" and
\verb"nCalls_2". The values are the same as in CalcHEP. Here is an
example:
{\footnotesize
\begin{verbatim}
nSess_1	:	5
nCalls_1	:	100000
nSess_2	:	5
nCalls_2	:	100000
\end{verbatim}
}
The defaults are the same as in CalcHEP.

\subsection*{Generator}
The following parameters of the event generation can be modified:
\begin{verbatim}
sub-cubes	:	1000
\end{verbatim}
The defaults are the \CalcHEP~defaults.

Examples of batch files and the output results
are given in section~\ref{sec-benchmarks}.

\subsection{Monitoring batch session}

The batch session is started with the command:
\begin{center}
\verb|./calchep_batch batch_file|
\end{center}
where \verb|batch_file| is the name of the file that contains the
batch instructions.  The batch program will print the following to
screen:
\begin{verbatim}
calchep_batch version vv

Processing batch:
Progress information can be found in the html directory.
Simply open the following link in your browser:
file:///WORK/html/index.html
You can also view textual progress reports in WORK/html/index.txt
	and the other .txt files in the html directory.
Events will be stored in the Events directory.
\end{verbatim}
where \verb|vv| is the version number and \verb|WORK| denotes the path
to the calchep working directory.
The user can view the progress in their favorite browser as well as
check the results and details.
Examples of the different batch files and the resulting output
are given in Section~\ref{sec-benchmarks}.
Among the details of the resutls, the html pages contain links to the
\verb|prt_1| and \verb|session.dat| files for each subprocess and each scan parameter.  These
files contain the full details of the VEGAS session including all the
parameters of the run.  At the bottom of the \verb|prt_1| file, there
is a history of the results of the VEGAS phase space integration.  The
user may find these useful.

\subsection{Results}
After the events and/or distributions are generated, they are stored
in the Events directory. The prefix of the files is the name specified
in the batch file plus either \verb|-single| if no scans were specified or
a string specifying the scan parameter values if one or more scans are
specified. We will assume this is filename in the following. If events
are requested, they will be stored in the files
\begin{verbatim}
filename.lhe.gz
filename.nt
\end{verbatim}
where \verb|filename.lhe| is the event file in Les Houches format and 
\verb|filename.nt| is in PAW ntuple format. The ntuple file is only
created if the keyword \verb"NTuple" is set to \verb|True| and
nt\_maker is present in the bin directory. If distributions are
requested, they will be stored in the files
{\footnotesize
\begin{verbatim}
filename.distr
filename_1.png
filename_2.png
...
\end{verbatim}
}
where \verb|filename.distr| is the raw distribution data and can be read by \verb"show_distr" in the bin directory. The distributions generated on the fly by the batch script are stored in the files ending in \verb".png".

\section{Particle interaction model implementation}
\label{models} 

A model of particle interaction in \CalcHEP~is stored  in five
tables, named  {\tt Parameters}, {\tt Constraints}, {\tt Particles}, {\tt Vertices} and {\tt Libraries}. These
tables are stored in the respective files {\tt varsN.mdl}, {\tt funcN.mdl}, {\tt prtclsN.mdl}, {\tt
lgrngN.mdl}, {\tt extlibN.mdl} which are located in {\tt WORK/models/}.
The {\tt N} in these file
names refers to the model number.
For all of these tables, a {\tt \%} at the beginning of any row means
that that row is a comment and \CalcHEP~ignores it.  We describe each
of these tables in this section.
  
\subsection{Independent parameters}\label{parameters}	      

The table {\tt Parameters} contains all the independent parameters of
the model.  It consists of three columns, for example,
\begin{center}
\begin{tabular}{|l|l|l|}
\hline
{\bf Name}    & {\bf Value}    & {\bf Comment}\\
\hline
EE  &0.3123      &elecromagnetic constant\\
SW  &0.481 & MS-BAR sine of the electroweak mixing angle\\
MZ         & 91.187       & Z-boson mass\\
Mh&  125          &higgs mass     \\
\hline
\end{tabular}
\end{center}
where
\begin{itemize}
\item [Name:] This is where the name of the parameter belongs.  
It can
contain up to 11 characters. The first character
must be a letter.  The others may be either letters or
digits. The underscore symbol is also permitted and 
\CalcHEP~is  sensitive to the case of the characters.
  There is a set of reserved names which cannot be used
for parameter names:
\begin{itemize}
\item {\tt i}    is reserved for the imaginary unit;
\item {\tt Sqrt2} is reserved for $\sqrt{2}$;
\item {\tt p1,p2,p3,\ldots} are reserved for particle momenta;
\item {\tt m1,\ldots,M1,\ldots} are reserved for Lorentz indices;
\item {\bf G5} is used for the $\gamma^5$ Dirac matrix;
\end{itemize}
There is another subtelty that should be considered when naming parameters.
Although \CalcHEP~is sensitive to the case of the parameters, Reduce
is not.  Therefore, if the user would like to use the \CalcHEP~results
 in Reduce, \heshe~ should distinguish all names by more than
case.  Additionally, although \CalcHEP~allows underscores as part of
parameter names, the underscore is treated differently by
Mathematica.  So, if the user would like to use the \CalcHEP~results
in Mathematica, \heshe~ should not use underscores in the parameter names.
    
\item [Value:] This is where the numerical value for the
  parameter is stored.  Dimensionful parameters should be in powers of
  GeV.

\item [Comment:] This is where the user can enter a description
  of the parameter.  It is ignored by \CalcHEP~and is purely for
  informational purposes.
\end{itemize} 

There are two parameters which are treated specially in \CalcHEP. 
They are \verb|GG| and \verb|Q|. The first is reserved for the strong 
coupling
$$  GG=\sqrt{4\pi\alpha_{qcd}} $$
\CalcHEP~forbids the appearence of this parameter in the  {\tt Constraint} table and in
the {\it Lorentz part} of the {\tt Vertices} table (see below). But,
it  can appear  
in the {\it Factor} column of the {\tt Vertices} table. This parameter is evaluated  by 
\CalcHEP~using  QCD parameters which the user can view and/or modify in Menu 4 of
Fig.\ref{num_module}.  \CalcHEP~ignores the value of \verb|GG| found
in the {\tt Paramters} table. The variable \verb|Q| is intented to be
used as a scale parameter in
the quark  effective masses and Yukawa couplings. When \CalcHEP~peforms a
calculation of the particle width, it sets  \verb|Q| equal to the
decaying particle's mass.  This scale is then used to calculate the
quark effectifve masses at that scale (see Sections \ref{sec:decaySLHA} and \ref{sec:slhaplus}).
In all other cases,  \verb|Q| is set equal to its value from the {\tt Paramters} table.

\subsection{Dependent parameters} 
	   \label{constraints}     
 The {\tt Constraints} table contains all the dependent parameters of
 the model.  It consists of two columns, for example, 
\begin{center}
{\footnotesize
\begin{tabular}{|l|l|}
\hline
{\bf Name} &{\bf Expression}\\
\hline
CW       & \verb|sqrt(1-SW^ 2)|                 \% cos of the Weinberg angle\\
MW       & \verb|MZ*CW|                         \% W-boson  mass  \\
GF       & \verb|EE^2/(2*SW*MW)^2/Sqrt2|        \% Fermi constant  \\
\%Local!    &\\
\hline
\end{tabular}
}
\end{center}
where
\begin{enumerate}
\item [Name:] This is where the name of the parameter belongs.
  The restrictions on the names are the same as for the independent
  parameter names.

\item [Express:] This is where the formula belongs which
  defines the value of this dependent parameter.  The formula can
  contain the following:
\begin{itemize}
\item integer and float point numbers, 
\item independent parameter names contained in the {\tt Parameters} table, 
\item dependent parameter names defined above the current row,
\item parentheses \verb|()| and arithmetic operators \verb|+, -, /, *, ^|,
\item the symbols \verb|i| and \verb|Sqrt2|,
\item standard functions from the \C~  mathematics library such as   
\verb|sqrt(x)| and \verb|sin(x)|.  The full list 
of these functions is contained in the {\tt \$CALCHEP/include/extern.h}
file,
\item functions from the SLHAplus package (see Section
\ref{sec:slhaplus}) which collects many
routines useful for model building,
\item the function \verb|if(x,y,z)| which returns \verb|y| if \verb|x|$>0$ and \verb|z| otherwise,
\item any user defined functions.  The code containing these functions
  should be included in the {\tt Libraries} table.  Their prototypes
  should also be included in the {\tt Libraries} table.  If their
  prototypes are not included, \CalcHEP~assumes they return double
  type.  A list of the resulting
auto-prototyped  functions  appears in the {\tt WORK/results/autoprot.h} file after compilation 
of the numerical code.
\end{itemize}
\end{enumerate}
Additionally, anything after the {\tt \%} symbol is considered a
comment and ignored.  This can be used to enter a comment about the
dependent parameters.  

Some models contain thousands of
dependent parameters.  For a particular process, only a small subset
of these is used. To reduce the
size of the generated code,  \CalcHEP~compiles only the 
dependent parameters   involved in the current matrix element calculation. 
However, it is sometimes necessary to have other dependent parameters
compiled with the matrix element code in order to apply constraints
such as on the particle spectrum.
For this reason, \CalcHEP~splits  all the dependent models parameters into two
sets: the {\it public} parameters and the {\it local} parameters. 
By default, the {\it public} parameters are those  that are required to
calculate all the particle masses and widths as well as those that
depend on external functions (except for dependence on standard \C~
math functions) and all dependent parameters above any of these.
All the dependent parameters below these are defined as {\it local}.
The {\it local} parameters are not compiled unless
they are needed for a matrix element calculation.
If the user would like to force \CalcHEP~to include a larger subset
of the dependent parameters in the numerical code, \heshe~ can place
the comment {\tt \%Local!} in the {\tt Constraints} table in the first
column.  \CalcHEP~will always include the parameters up to, at least,
this point.

The {\tt Constraints} menu item of Fig.~\ref{num_module} displays only
{\it public} parameters.  These can be used along with { \tt global} parameters
in the  specification  of  cuts and histogram  limits,  user-defined kinematical
functions, and the QCD scale  (see Section \ref{sec: numerical}).
The user-defined external functions can get the  values of the  independent 
model  parameters and  {\tt public} dependent model parameters with
the routine
\begin{verbatim}
   int findval(char *name, double *value); 
\end{verbatim}
where \verb|name| is the parameter name and \verb|*value| is filled
with the value of the parameter.  More details can be found in Section \ref{setting_parameters}.

\subsection{Particles} \label{particles}
The particles are defined in the {\tt Particles} table which consists
of 11 columns.  Each particle anti-particle pair is described by one
row of the table. For example
\begin{center}
\begin{tabular}{|l|c|c|c|c|c|c|c|c|c|c|}
 \hline
 Full  name  & A  & A  &  PDG   &2*spin& mass &width &color&aux& LaTex(A) & LaTeX(A+) \\
\hline
W-boson      &W+  &W-  &24      &2     &MW    &wW    &1    &G  &\verb|W^+| &\verb| W^-|\\
Higgs        &h   &h   &25      &0     &Mh    &!wh   &1    &   &h &h\\
\hline
\end{tabular}
\end{center}
The columns are:
\begin{itemize} 
\item[Full name:]  The full name of the particle can be
  entered here.  It is not used directly by \CalcHEP.  It is used to
  clarify what the short particle names mean.
\item[A\&Ac:]  These columns are where the
  particle name and antiparticle name belong.  More precisely, these
  columns contain the quantum field and its C-conjugate.  The field
  operator acting on the vacuum is understood to create the
  corresponding anti-particle.  
Self-conjugate fields
  (such as photons and Majorana neutrinos) should contain the same
  name in both columns.  
Any printing character can be used in the particle name except white
space, parentheses and the percent symbol (\%).  The length of the
particle name can not exceed 8 symbols.
For long particle names, we
should note that the graphical representation of the diagrams might
contain overlapping symbols.  

\item[PDG:] This is where the PDG code \cite{ParticlesFields}
  belongs.  This number is used mainly for interfacing with other
  packages.  For example, these codes are included in event files
  \cite{Alwall:2006yp}  in
  order to communicate the particle flavor to other programs.  
The parton distribution functions are also applied according to this
number.  The conventional PDG codes should be used for SM particles.
For other particles, the user should ensure that the code is not
reserved for another particles such as a meson or baryon.  Otherwise,
conflicts could arise when passing events to other programs, such as Pythia.

\item[2*Spin:] This is where the spin of the particle is
  specified.  It should be entered as the integer equal to twice the
  spin.  In other words, {\tt 0} should be entered for a scalar field,
  {\tt 1} for a spin $1/2$ fermion, {\tt 2} for a vector boson, {\tt
    3} for a spin $3/2$ fermion and {\tt 4} for a spin-2 boson.  Spin
  $3/2$ and $2$ particles should be massive.

\item[Mass:] This is where the mass of the particle is
  entered.   If massless, {\tt 0} can be entered.  Otherwise, it must
  be a parameter name which is defined in either the {\tt Parameters}
  table or the {\tt Constraints} table.

\item[Width:] This is where the width of the particle is
  entered.  If the particle is stable, {\tt 0} can be entered.
  Otherwise, it must be a parameter name.  In this case, however,
  this parameter can be defined in the {\tt Parameters} table, the
  {\tt Constraints} table or it can be preceeded with the {\tt !} symbol.  If
  it is preceeded with the {\tt !} symbol, \CalcHEP~will
  automatically calculate it when needed.  In this case, the parameter
  should not appear in either the {\tt Parameters} table or the {\tt
    Constraints} table.  When automatically calculating the width,
  \CalcHEP~first attempts the $1\to2$ decays.  If none are found, it
  attempts the $1\to3$ decays.  If none are still found, it attempts
  the $1\to4$ decays.  If none are found at this point, it takes the
  width as zero.  If information about particle widths was read from a
  SLHA file (See section  \ref{sec:slhaplus}), then the  widths are not
  evaluated  and \CalcHEP~substitutes the values it read from the SLHA
  file in the  propagators.

\item[Color:] This is where the color (SU(3)) representation
  is specified.  Supported representations are the singlet (specified
  by a {\tt 1}), the fundamental triplet (specified by a {\tt 3}) and
  the octet (specified by a {\tt 8}.)  If the particle is specified as
  a triplet, the antiparticle is treated as an anti-triplet (the
  $\bar{3}$ representation.)

\item[Aux:] This field is primarily used to modify the propagator of
   the field.  For most fields, this column will be left blank.  The
   other possibilities are:
\begin{itemize}
\item  {\tt * :} specifies that the propagator should be point like
  (all momentum dependence is dropped.)
  This can be used to construct 4-fermion propagators, for example.
   These fields can not appear as external states of processes.  
\item {\tt l} and {\tt r :} are used to specify that a fermion is
  purely left and right handed respectively.  This can only be applied
  to massless fermions.  The effect of this is that when \CalcHEP~
  averages over the spin of the incoming fermion, it takes into
  account that there is only one polarization for this particle.  This
  is used, for example, for the SM neutrinos.
\item {\tt g :} declares that the vector particle is treated as a {\it gauge}
boson. In this case t'Hooft-Feynman gauge is used for the vector boson
propagator and the ghost
fields  {\tt A.c} and {\tt A.C} (where {\tt A} is the name of the
vector boson) as well as the Goldstone boson {\tt
  A.f} can contribute to the Lagrangian. 
A massless vertor particle must be treated as a {\it gauge} boson.
In the absence of {\tt g} in this column, the unitary gauge is always used
for massive vector bosons and the ghosts and Goldstones associated
with it are not used in Feynman diagrams.
\end{itemize} 
The Formulaes for the particle's propagators are presented in Section
(\ref{Propagators}).

The {\tt Aux} column can also be used to specify a particle's electric
charge.
This charge is required by many external packages.  \CalcHEP~already
knows the charge of the SM particles and assigns it according to the
particle's PDG code.  It can determine the charges of many BSM
particles by analyzing the Feynman rules and assuming they conserve
electric charge.  However, for some particles, this will not be
sufficient to determine their charge.  For this reason, \CalcHEP~
allows to specify the charge of any BSM particle.  Specifically, three
times the charge should be entered in the {\tt Aux} field.  For
example, a particle with electric charge of $-1$ would be enetered as
{\tt -3}, a particle with electric charge of $2/3$ would be entered as
{\tt 2} and so on.  This charge must be written before other symbols
in this column (if there are any.)  We reiterate that this charge is not used to
define the Feynman rules of the photon in calculations done by
\CalcHEP~.  The interactions of the photon are entered in the {\tt
  Vertices} table along with all other Feynman rules (see
Subsection~\ref{lagrangian}.)  The electric
charge defined in the {\tt Aux} column of the {\tt Particles} table is
only used to communicate with other programs that require
it.  

\item[LaTeX:]  This is where the
  \LaTeX~ symbol for the particle and antiparticle are entered.  These
 symbols are used when \CalcHEP~produces \LaTeX~ output for the
 Feynman diagrams that it constructs.  
\end{itemize}
	
\subsection{Interaction vertices} \label{lagrangian}	      

   The {\tt Vertices} table contains the Feynman rules for the model. For
example, 
\begin{center}
\begin{tabular}{|l|l|l|l|l|l|}
\hline
A1   &A2   &A3   &A4   &         Factor               &  Lorentz part\\
\hline
h    &W+   &W-   &     &EE*MW/SW                      &m2.m3\\
\hline
\end{tabular}
\end{center}

   The first four columns ({\tt A1}, {\tt A2}, {\tt A3} and {\tt A4})
   specify the particles[antiparticles] involved in the
   interaction.  These must be the particle names
   defined in the {\tt Particles} table.  The last of these {\tt A4}
   may be empty, which specifies a three-point vertex.  The first
   three columns must be nonempty.  (The propagators are not specified
   in this table.  They are hard coded.  Section \ref{Propagators} contains
   further details.)

   The last two columns, {\tt Factor} and {\tt LorentzPart} define the
   vertex.  If $S$ is the action for a particular vertex, the vertex
   can be obtained by functionally differentiating with respect to the
   fields in the vertex as in
\begin{equation}
\label{funcDeriv}
\frac{\delta S}{\delta A1_{[m_1]}(p_1)\, \delta A2_{[m_2]}(p_2)\, 
\delta A3_{[m_3]}(p_3)\,
 [\delta A4_{[m_4]}(p_4)]}
  =  \hspace{1in}
\end{equation}
\begin{displaymath}
(2 \pi)^4\delta^4(p_1+p_2+p_3\, [+p_4]) \, [{C^{-1}}^T] \, ColorStructure \cdot 
 Factor \cdot LorentzPart\;,
\end{displaymath}   
where $p_i$ and $m_i$ refer to the 4-moment and Lorentz indices (if
any.)  The square brackets ($[$ and $]$) denote parts of the
expression which only appear in some vertices, but not others.  
The Fourier transform  is defined as
\begin{equation}
       A(x) =  \int \frac{d^4k}{\left(2\pi\right)^4} \ e^{-i k\cdot x}  A(k)\;.
\label{Fourier}
\end{equation}
The other pieces of Equation (\ref{funcDeriv}) will be discussed below.

The {\tt Factor} column is where the $Factor$ from Equation
(\ref{funcDeriv}) belongs.  This must be a rational  monomial
constructed from the model parameters, integer numbers and the
imaginary unit ({\tt i}).  It is best to factor as much as possible
from the $LorentzPart$ since the $LorentzPart$ of the Feynman diagrams
is usually the most time consuming and memory intensive part of the
calculation.  

The   {\tt LorentzPart} column is where the $LorentzPart$ from
Equation (\ref{funcDeriv}) belongs.  It  must be  a Lorentz tensor or a  Dirac $\gamma$-matrix
expression. The coefficients of the terms in this expression can be
polynomials of the model 
parameters and scalar products of the momenta.  The division operator
({\tt /}) 
is forbidden from this column.  It must be transferred to the {\tt
  Factor} column or into a model parameter.

The notation for Lorentz indices, momenta, contractions,
and the metric tensor are similar to those in the \REDUCE~ package. 
The Lorentz indices of the fields in the vertex are labeled by a {\tt
  m} for the first index and a {\tt M} for the second index followed
by the particle number for that vertex.  For example, a vector field
in the third column would have Lorentz index {\tt m3} while a tensor
field in the second column would have Lorentz indices {\tt m2} and
{\tt M2}.  The momenta use the symbol {\tt p}
followed by the same number.  For example, a scalar field in column 1
would have momentum {\tt p1}.
 A dot
({\tt .}) is placed between two momenta, a momentum and its Lorentz
index, and between two Lorentz indices (for the metric tensor.)  Here
are some examples:
\begin{center}
\begin{tabular}{lcl}
 {\tt p1.p2} & means & $p_{1\mu}\ p_2^\mu$ \\
 {\tt p1.M2} & means & $p_1^{M_2}$\\
 {\tt m1.m2} & means & $g_{m_1 m_2}$.  
\end{tabular}
\end{center}

Dirac $\gamma$-matrices are written with a {\tt G} and the momentum or
Lorentz index in parentheses, while the $\gamma_5$ matrix has a
{\tt 5} without parentheses.  For example, we have:
\begin{center}
\begin{tabular}{lcl}
{\tt G(m1)} & means & $\gamma^{m_1}$\\
{\tt G(p2)} & means & $\gamma^\mu p_{2\mu}$\\
{\tt G5} & means & $\gamma_5$
\end{tabular}
\end{center}
The $\gamma_5$ matrix  is defined by 
   $$\gamma_5 = i\; \gamma_0\gamma_1\gamma_2\gamma_3\ .$$
The anti-commutation relation for the gamma matrices in \CalcHEP~
notation is
\begin{center}
\verb|G(v1) G(v2) + G(v2) G(v1) = 2 v1.v2|\ ,
\end{center}
where  \verb|v1| and \verb|v2| are either momenta or Lorentz indices.

In the case of anti-commuting fields the functional derivative in
Equation~(\ref{funcDeriv}) is assumed to act from the right.
The number of fermion fields in a vertex must be either two or zero.
If the user would like to implement a four-fermion interaction,
\heshe~ must use an unphysical auxiliary field with a point-like
propagator (see Subsections \ref{particles} and \ref{Propagators} for further details.)

\CalcHEP~interprets the  anti-particle spinor  field as a  
C-conjugated particle field, rather than the Dirac conjugated field.
These definitions are related to each other by
\begin{equation}
 \frac{\delta}{\delta \psi^c}={C^{-1}}^T\frac{\delta}{\delta\bar{ \psi}}
\label{C_appearence}
\end{equation}  
which is the reason for the appearance of the ${C^{-1}}^T$ matrix
in Eq. (\ref{funcDeriv}).
The particle and anti-particle fields can appear in the  vertices in any order.
Vertices can also contain two particle fields or two antiparticle
fields.  In other words, vertices that violate fermion number are allowed.

Any fermion vertex
can be written in two forms which depend on the order of the fermion fields. 
After permutation of the fermion fields, the  {\tt LorentzPart} is transformed 
according to  
\begin{equation}
  G(v_1) \, G(v_2)  \ldots [G5] \ldots  G(v_n)  \rightarrow 
  (-G(v_n)) \ldots [G5] \ldots  (-G(v_2)) \, (-G(v_1)) \;,
\label{reverse}
\end{equation}
where the order of the gamma matrices is reversed and each gamma
matrix with a Lorentz index gets a sign change while the $\gamma_5$
matrix does not get a sign change.

We note that the definition in Eq. (\ref{funcDeriv}), the $LorentzPart$ 
has the appropriate symmetry property when
identical particles appear in the vertex. This symmetry is not checked by
\CalcHEP, and its absence will lead to the wrong results.
Equation \ref{reverse}  can  be used to check this symmetry in the case of 
two identical Majorana fields in one vertex.  It should also be noted
that in the case of $n$ identical particles, the functional derivative (\ref{funcDeriv})
gets a corresponding  factor of $n!$ which should be included in the vertex.

The totally antisymmetric Levi-Civita tensor can be used in vertices.
It is given by \verb|eps(v1,v2,v3,v4)|, where \verb|v1|, \verb|v2|,
\verb|v3|, and \verb|v4| are either momenta or Lorentz indices.

The $ColorStructure$ from Eq. (\ref{funcDeriv}) is not included in the
{\tt Vertices} table.  \CalcHEP~substitutes it in automatically
according to the following rules:  If all the particles in the vertex
are color singlets, \CalcHEP~inserts \verb|1|.  If the vertex
contains one fundamental and one antifundamental ($3\times\bar{3}$),
the identity matrix is inserted.  If the vertex contains two color
octet fields ($8\times8$), the identity matrix is inserted.  If the
vertex contains three color octet fields ($8\times8\times8$), it
inserts
\begin{displaymath}
-i f(a1,a2,a3)
\end{displaymath}
where $f^{a1}_{a2,a3}$ is the structure constant of SU(3) and
the color adjoint indices $a1, a2$, and $a3$ are taken in the same order
they appear in the {\tt Vertices} table.  If the vertex contains a
fundamental, an antifundamental, and a color adjoint field
($3\times\bar{3}\times8$), \CalcHEP~inserts
      $$\frac{1}{2}\lambda(\bar{i},j,a),$$
where $\lambda(\bar{i},j,a)$ are the Gell-Mann matrices.
Other color structures are not implemented in \CalcHEP, however, it is
possible to construct them by means of an unphysical auxiliary field
(see Subsections \ref{particles}, \ref{Propagators} and \ref{tensor_propagator} for further details.)

\subsection{External functions and libraries.}\label{extlib}         
The {\tt Libraries} table is used to link external code and declare
external functions.
\begin{center}
\begin{tabular}{|l|}
\hline
External libraries  \\
\hline
\verb|%  LHAPDF shared library linking  |\\
\verb|-L $(HOME)/Packages/lhapdf-5.8.4/install/lib -lLHAPDF|\\
\verb|%  LHAPDF static library linking |\\
\verb|% $(HOME)/Packages/lhapdf-5.8.4/install/lib/libLHAPDF.a  -lgfortran|\\
\verb|% function declaration|\\
\verb|extern  double bsgnlo(void);|\\
\verb|%  user library |\\
\verb|   $(HOME)/my_code/bsgnlo.a |\\ 
\hline
\end{tabular}
\end{center}

  Lines beginning
with a \verb|%| are comments and are ignored by \CalcHEP.  Lines
beginning with the keyword \verb|extern| are considered to be
prototypes of external functions defined in external code.  These
lines should include the full function prototype (including the
semicolon at the end) on one line in the syntax of
the \C~ programming language.  These functions can be used in the
definitions of the dependent parameters in the {\tt Constraints}
table (see Subsection~\ref{constraints}.)  

External code and libraries can be linked to the numerical code by
using this table as well.  The user should enter a list of the
external code, libraries and any flags necessary for \hisher~ model in
this table.   Some typical examples of external code are user defined
kinematical variables which can be used in cuts and histograms (see
Subsection~\ref{functions}) and the LHAPDF libraries (see
Subsection~\ref{sec:LHAPDF_linking}.)  All lines which do not start with \verb|%|
or \verb|extern| are concatenated and passed to the linker which
creates the executable for numerical calculations.  These lines can
make use of environment variables.  \CalcHEP~defines two in it's
startup scripts (\verb|calchep| and \verb|calchep_batch|) that the
user can make use of.  They are \verb|$CALCHEP| which is the path to
the \CalcHEP~root directory and \verb|$WORK| which is the path to the
user's working directory.  The user can also make use of \hisher~own
environment variables.  These environment variables can be used with
or without parentheses (either \verb|$CALCHEP| or \verb|$(CALCHEP)| is
acceptable.)  \CalcHEP~will translate between the two depending on
whether they are used in a {\tt Makefile} or in a shell environment.

For functions presented in the SLHAplus package (Section \ref{sec:slhaplus})
prototyping and special link instructions are not needed.

\subsection{Propagators} \label{Propagators} 
\CalcHEP~defines the propagators for particles of spin less than or
equal to two.  These propagators are hard coded and not modifiable 
by the user unless specified below.  \\
{\tt Spin 0:}  The spin-0 propagator is given by
      $$ <0|T[A(p_1) , A^+(p_2)]|0> = \Delta_c(p_1,p_2,M) = \frac{i\delta(p_1 + p_2)}
{(2\pi)^4(p_1^2 - M^2)}\;\;.$$
{\tt Spin 1/2:}  The spin-1/2 propagator is given by
      $$<0|T[A(p_1)  , \bar{A}(p_2)  ]|0> =  (\not p_1 + M) \,
\Delta_c(p_1,p_2,M)\;\;\;,$$
where $\not p = p^{\mu}\gamma_{\mu}$.
If the fermion is defined to be purely left or right handed (see Subsection~\ref{particles}), the
propagator is defined as
         $$\frac{  \not p_1  (1 \pm \gamma_5)}{2} \; \Delta_c(p_1,p_2,M)\;\;. $$
{\tt Spin 1:}  In unitary gauge, the propagator is given by
\begin{equation}
  <0|T[A^{m_1}(p_1),  \; (A^{m_2})^+(p_2)]0>  =  -(g^{m_1 m_2} + \frac{p_1^{m_1}
p_2^{m_2}}{M^2})
\Delta_c(p_1,p_2,M),
\label{mvPropagator}
\end{equation}
while in t'Hooft-Feynman gauge, it is given by                 
              $$ -g^{m_1m_2} \, \Delta_c(p_1,p_2,M)\;.$$
We remind the user that a massless vector particle must be defined as
a gauge boson (see Subsection~\ref{particles}.)\\
{\tt Spin 3/2:}   The spin-3/2 propagator is given by
\begin{eqnarray}
<0|T[A^{m_1}(p)  , \bar{A}^{m_2}(p')]|0> =\left(-3(\not{p}+M) (g^{m_1 m_2}
-\frac{p^{m_1}p^{m_2}}{M^2})\right.  \\
\nonumber
- \left.(\gamma^{m_1}+\frac{p^{m_1}}{M})(\not{p}-M)(\gamma^{m_2}+\frac{p^{m_2}}{M})\right)\Delta_c(p,p',M)
\nonumber
\end{eqnarray}
{\tt Spin 2:} The spin-2 propagator is given by
\begin{eqnarray}
\nonumber  
 <0|T[A^{m_1 \mu_1}(p),(A^{m_2 \mu_2})^+(p')]0>=\Bigg((g^{m_1 m_2}+2\frac{p^{m_1}p^{m_2}}{M^2})(g^{\mu_1 \mu_2}
+2\frac{p^{\mu_1}p^{\mu_2}}{M^2}) \\
\nonumber
  -3(g^{m_1 \mu_1}p^{m_2}p^{\mu_2} +g^{m_2 \mu_2}p^{m_1}p^{\mu_1} +g^{m_1\mu_2}p^{\mu_2}p^{\mu_1}+g^{m_2.\mu_1}p^{m_1}p^{\mu_2}
  )M^{-2}\\
\nonumber
   +3(g^{m_1 \mu_1}g^{m_2 \mu_2}  + g^{m_1 \mu_2} g^{m_2 \mu_1} -g^{m_1 m_2} g^{\mu_1
\mu_2})\Bigg) \Delta_c(p,p',M)
\nonumber
\end{eqnarray}
{\tt Auxiliary propagators:}  When massive particles are marked as auxiliary
fields (see Subsection~\ref{particles}) by putting a \verb|*| in the
\verb|Aux| column, the momentum dependence of the propagator
is removed.  
$\Delta_c(p_1,p_2,M)$ is replaced with
$$ \frac{\delta(p_1+p_2)}{ (2 \pi)^4 i \;M^2}\;$$
and all terms proportional to the particle momentum $p$ 
in the   numerator are dropped. Auxiliary particles cannot appear as
incoming or outgoing states.  They are only used to implement  point-like
interactions.

\subsection{Ghost and Goldstone  propagators} \label{ghostFields} 
In addition to the fields enumerated
in the {\it Particles} table, the Lagrangian can depend on a few other fields.
In particular, gauge theories have Faddeev-Popov ghosts \cite{BD} and, if
broken, Goldstone bosons.  Furthermore, complex color structures
require a special tensor auxiliary field.  All of these fields are
automatically generated by \CalcHEP~where appropriate by adding a
final \verb|.c|, \verb|.C|, \verb|.f|, \verb|.t| or \verb|.T| as described below.

\paragraph{ Faddeev-Popov ghosts and anti-ghosts}  are  generated for
any gauge vector particle which is  marked by a {\tt g} in the {\it 'Aux'} column 
of the {\it Particles} table (see Subsection~\ref{particles}.)  The
names of the Faddeev-Popov ghosts and anti-ghosts are constructed by
adding a \verb|.c| and \verb|.C|, respectively, to the particle name.
For example, if the gluon is named \verb|G|, the gluonic ghost is
named \verb|G.c| and the gluonic anti-ghost is named \verb|G.C|.  The
ghosts and anti-ghosts corresponding with the \verb|W+| and \verb|W-|
gauge bosons are \verb|W+.c|, \verb|W+.C|, \verb|W-.c| and \verb|W-.C|.
Hermitian  conjugation  transforms a  Faddeev-Popov ghost
into a ghost with the same sign whereas it changes the sign of the anti-ghost.  For example,
\begin{eqnarray*}
(G.c)^+&=&G.c\\  
(G.C)^+&=&-G.C\\
 (\mbox{W+.c})^+&=& \mbox{W-.c} \\  
(\mbox{W+.C})^+&=&-\mbox{W-.C}
\end{eqnarray*}
Faddeev-Popov (anti)ghosts are anti-commuting, scalar
fields\footnote{The
 well-known spin-statistics relation is not valid for unphysical
fields.}.  
 The nonzero propagators for these fields are:
$$<0|T[\mbox{A+.c}(p_1), \; A.C(p_2)]|0> = <0|T[\mbox{A+.C}(p_1), \;
A.c(p_2)]|0>  = \Delta_c(p_1,p_2,M),$$
where $A+$ is the conjugate of $A$ and  $M$ is the mass of the parent particle (we are assuming Feynman
gauge.)

The reason \CalcHEP~introduces the Faddeev-Popov ghosts at tree-level
is that it sums over the unphysical polarizations of the gauge bosons in
the external states as well as the physical polarizations 
in order to reduce precision loss due to
large cancellations.  The Faddeev-Popov ghosts (and the Goldstone
bosons for a broken gauge theory) are required to cancel the
unphysical polarizations.  (See \cite{BD} for further details.)

\paragraph{Goldstone bosons}  are related to broken symmetries.  In
the case of broken gauge symmetries, they become the longitudinal
degrees of freedom of the gauge boson.  \CalcHEP~automatically
generates these fields for massive vector bosons by appending a
\verb|.f| to the end of the gauge boson name.  For example, the
\verb|W+| and \verb|W-|
gauge bosons have the Goldstone bosons \verb|W+.f| and \verb|W-.f|
associated with them.  These Goldstone bosons are commuting, scalar
fields that satisfy the same conjugation rules as the gauge boson they
belong with.  For example, $(W+.f)^+= W-.f$.
The nonzero propagators for these fields are:
$$T[\mbox{A+.f}(p_1), \; A.f(p_2)]   = \Delta_c(p_1,p_2,M),$$
where $A+$ is the conjugate of $A$ and  $M$ is the mass of the gauge boson (again we consider Feynman
gauge.)

\paragraph{ Auxiliary tensor field.}  

  Whereas  the  Faddeev-Popov ghosts  and Goldstone bosons are standard
elements of modern quantum field theory, this auxiliary tensor field
was invented by the original \CalcHEP~authors in order to construct
complicated color vertices such as the four-gluon vertex.  These
auxiliary fields are automatically generated whenever a particle is
defined with a nontrivial $SU(3)$ color representation by adding
\verb|.t| and \verb|.T| to the particle name. Two auxiliary tensor
fields are generated automatically and are typically used for a
constraint and a Lagrange multiplier.  

These auxiliary fields are commutative and satisfy the same
conjugation rule as the parent particle, while it is  Lorentz-transformed 
like  a tensor field. The propagator is point like
\begin{equation}
<0|T[\mbox{A+.t}^{m_1M_1}(p_1), \; A.t^{m_2M_2}(p_2)]|0> = \frac{1}{(2 \pi)^4i} \; 
\delta(p_1+p_2) \, g^{m_1m_2}\, g^{M_1M_2}\;.
\label{tensor_propagator}
\end{equation}

Further information about the use of the Faddeev-Popov ghosts,
Goldstone bosons and auxiliary tensor fields can be found in 
the manual.
            

Although it is possible to implement a new model of particle
interactions directly using the table definitions described here, for
complicated models with a large number of particles, parameters
and Feynman rules, it is a good idea to use an external program to
generate the model files.  In section \ref{modelGeneration}  we briefly describe
available packages  for model implementation.

\section{\label{sec: tools}Tools for model implementation and the model
  repository at HEPMDB}

\subsection{The SLHAplus package}\label{sec:slhaplus}

The SLHAplus\cite{Belanger:2010st}  package  included in 
\CalcHEP~supports the reading of SLHA formatted parameter files.  It
also allows the diagonalization of mass matrices on the fly.  
In many models of elementary particles, there are significant loop
corrections to  particle masses. Several packages have been created which 
perform these loop calculations for MSSM-like models. For example,
Isajet\cite{ISAJET},  SoftSusy\cite{Allanach:2001kg}, Spheno\cite{Porod:2003um},
SuSpect\cite{Djouadi:2002ze}, and
NMSSMTools\cite{Ellwanger:2006rn}. An agreement has been formed to
pass the input
parameters and  the calculated 
particles spectra and mixing matrices  via text files in the special format {\tt
SLHA}\cite{Skands:2003cj,Allanach:2008qq}.  To read SLHA  file, the user can enter the following command in the Constraints table
\begin{verbatim}
rd  |slhaRead( fileName ,mode)
\end{verbatim}
The second argument of \verb|slhaRead| specifies the kind of  data to
read. Generally,  $mode=m_1+2m_2+4m_4+8 m_8$ where:
\begin{center}
\begin{tabular}{lll}
 {\tt m1} & $0/1$ &keep new data /  keep old data\\
 {\tt m2} & $0/1$ &ignore mistakes /   stop if a mistake in the input file is encountered\\
 {\tt m4} & $0/1$ &read DECAY     /   don't read   DECAY \\
 {\tt m8} & $0/1$ &read BLOCK     /   don't read   BLOCK \\
\end{tabular}
\end{center}
To get the values of parameters after reading the \verb|slhaVal|
function can be used as in:\\
\verb|   Zt12 | slhaVal("STOPmix",MZ,2,1,2)   
where the first argument of \verb|slhaVal| is the name of the BLOCK to
read, the second argument is the scale
 (in this example the scale parameter is not specified in the BLOCK so
 this is not used), the third parameter fixes the number
of parameters for each line of the BLOCK, and the parameters follow.

 A SLHA file also can contain information about
particle widths and decays channels.   If such information is read
(m4=0), \CalcHEP~uses the values read from the SLHA file in place of
an automatic calculation if the \verb|!| was used in the particle table.

The SLHAplus package  in CalcHEP is considered as a collection of tools
for realization of {\it Constraints}.  Except of
SLHA interface it contains routines for matrix diagonalizing  adapted for
direct implementation in CalcHEP tables and  supports an implementation of the QCD running
couplings\\
\verb|     McRun(Q) , MbRun(Q), MtRun(Q)| \\
and effective quark masses for Yukawa couplings needed for Higgs width
calculation\\
\verb|     McEff(Q),  MbEff(Q), MtEff(Q)|\\
See detals and examples in \cite{Belanger:2010st}.

\subsection{\label{sec:HAA HGG}Effective Higgs $\gamma$-$\gamma$  and {\it glue}-{\it glue}
interactions}  

\subsubsection*{Construction of effective vertices}
The interactions of the Higgs field with electrically charged and
colored particles induce interactions with the gluon ({\it h-glu-glu})
and photon ({\it
  h}-$\gamma$-$\gamma$) at
loop order.  These interactions lead to important contributions to the
Higgs production and decay.  
The leading induced effective operators for these interactions can be written as
\cite{Djouadi:2005gi, Djouadi:2005gj}: 
\begin{eqnarray}  
\label{hff}
h\bar{\psi} \psi           & \to & \frac{\alpha}{8\pi}   f^c_{\psi} q_{\psi}^2 h F^{\mu\nu}F_{\mu\nu}            A_{1/2}    (\frac{{M_h}^2}{4M_{\psi}^2})/M_{\psi} \\
\label{hvv}   
M_v h \bar{v}_{\mu}v^{\mu} & \to & \frac{-\alpha}{16\pi} f^c_{v}    q_{v}^2    h F^{\mu\nu}F_{\mu\nu}            A_{1}      (\frac{{M_h}^2}{4M_{v}^2})   /M_{v}\\
\label{hss}
M_s h \bar{s}s             & \to & \frac{\alpha}{16\pi}  f^c_{s}    q_{s}^2    h F^{\mu\nu}F_{\mu\nu}            A_{0}      (\frac{{M_h}^2}{4M_{s}^2})   /M_{s}\\
\label{h5ff}
h\bar{\psi}i\gamma_5 \psi  & \to & \frac{\alpha}{16\pi}  f^c_{\psi} q_{\psi}^2 h F^{\mu\nu}\tilde{F}_{\mu\nu}\tilde{A}_{1/2}(\frac{{M_h}^2}{4M_{\psi}^2})/M_{\psi}
\end{eqnarray}
where $\alpha$ is either the strong or electromagnetic coupling, $f^c$ is a color
factor which depends on the color representation of the virtual
particle.  For a fundamental SU(3) representation, $f^c=3$ for photon vertices and 
$f^c=1/2$ for gluon vertices. For an adjoint representation, $f^c=8$
for a photon and $f^c=-2$ for a gluon.
$q$ is the  electric charge of the particle in the loop for the
$\gamma\gamma$ operator and $1$ for the gluon operator.
The functions $A_{1/2}$,  $A_{1} A_{0}$,  $\tilde{A}_{1/2}$ can be
found in
\cite{Djouadi:2005gj}.  They are implemented in the SLHAplus package
  and have the names {\tt HggF, HggV, HggS, HggA}, respectively.
These functions
return a complex value where the imaginary part appears if the argument is larger than
one. 

In \CalcHEP~notation, the $\lambda hF_{\mu\nu}(A)F^{\mu\nu}(A)$ vertex
is implemented as
\begin{verbatim}
P1 |P2 |P3 |P4 |>   Factor <|>dLagrangian/dA(p1)dA(p2)dA(p3)                                                                <|
A  |A  |h  |   | -4*lambda  |(p1.p2*m1.m2-p1.m2*p2.m1)
\end{verbatim}
whereas the TP-odd interaction
 $\lambda hF_{\mu\nu}(A)\tilde{F}^{\mu\nu}(A)$ is written as 
\begin{verbatim}
A  |A  |h  |   | -4*lambda    |eps(p1,m1,p2,m2)
\end{verbatim}

\CalcHEP~assumes that these vertices are self-conjugate, but 
in general, the 
hVV couplings are complex. However, we note that after squaring and
summing over the diagrams, the result is the same as if we replace the
complex coupling with its
absolute value after summing all the amplitudes.  
That is, the complex couplings from all the effective operators should
be summed.  The value of \verb|lambda| should be set equal to the
absolute value of this sum.

\subsubsection*{QCD corrections to $h\gamma\gamma$ coupling}
There are important QCD corrections for the  $h\gamma\gamma$  effective
vertex induced by colored particles in the loop.  
This correction can be described by an overall factor for the coupling 
\begin{equation}
        1+\frac{\alpha_s}{\pi}C_l( \frac{{M_h}^2}{4M_{l}^2})
\end{equation} 
where $M_l$ is the mass of the loop particle. The $C_l$ functions are known  for 
vector, axial-vector, and scalar interactions if the loop particle has color 
dimension 3.    The formulas  are 
rather cumbersome and only have simple analytic forms in asymptotic limits. 
The SLHAplus package  contains the {\tt  HgamF($\tau$), HgamA($\tau$), HgamS($\tau$)}
complex functions which interpolate between  tabulated data for the $C_l$
functions for fermion loops with scalar interaction, fermion loops with
pseudoscalar interactions, and scalar loops, respectively. The HDECAY
package was used to generation the tables. The appropriate QCD scale for
$\alpha_s$ is $M_h/2$. It allows to effectively take
into account large logarithmic corrections  at higher order
\cite{Djouadi:2005gi}.

\subsubsection*{QCD corrections for {\it hgg} coupling}

The $h\to GG$  process at NLO contains  radiative corrections which
are plagued by infrared divergences which cancel against infrared
divergences in loop diagrams caused by virtual gluons attached to external
legs. For this reason, QCD NLO corrections are presented for partial
widths and cross sections, but not for effective vertices as in the
$h\gamma\gamma$ case. \CalcHEP~users have to implement these NLO factors to vertices in 
form 
$$ \sqrt{ 1+ \frac{\alpha_s}{\pi}F_{nlo}} $$
where $F_{nlo}$  is the NLO contribution for the $hGG$ partial  width.

The QCD correction  for the $hGG$  interation induced by heavy quark  loops are
known at NNLO.  In case of a scalar (\ref{hff}) and preudo-scalar
(\ref{h5ff}) interaction 
they are, respectively \cite{Chetyrkin:1997iv}, \cite{Chetyrkin:1998mw}
\begin{equation}
\label{hggNNLO}
F_{nlo}^{\bar{q}q} = \left(\frac{95}{4} -\frac{7n_f}{6}\right)   
   + \frac{\alpha_s}{\pi}\left( 370.20  -47.19 n_f +0.9018 n_f^2
-(\frac{19}{8}+\frac{2n_f}{3})\log{\frac{M_f^2}{M_h^2}}\right)
\end{equation}
\begin{equation}
\label{hggNNLO5}
F_{nlo}^{\bar{q}\gamma_5 q} =
 \frac{221}{12}    
   + \frac{\alpha_s}{\pi}\left(171.5  -5\log{\frac{M_f^2}{M_h^2}}\right)
\end{equation}
The last expression is presented for $n_f=5$. The quark masses in
these expressions should be the pole values 
\cite{Spira:1995rr} and $\alpha_s$ should be calculated at $M_h$ scale.
The $\alpha_s^3$   QCD corrections to the hgg vertex in case of a
massive fermion  loop  when $M_f \gg Mh/2$  
was calculated in \cite{Baikov:2006ch} and appears to be at about the
$\approx 1\%$ level. 

The NLO QCD corrections for effective vertices induced by loop of scalar color
particle ( for instance SUSY {\it squark})  is
known at NLO and appears to be a factor of  $17/6$  larger than
$F_{nlo}^{\bar{q}q}$ \cite{Dawson:1996xz}
\begin{equation}
\label{hggNLOS}
F_{nlo}^{ss} = \left(\frac{319}{12} -\frac{7n_f}{6}\right) 
\end{equation}

Equations (\ref{hggNNLO}, \ref{hggNNLO5}, \ref{hggNLOS}) were obtained
in the limit of  
$ M_{f,s}<< M_h/2$. 
However,  it  appears that they work well even beyond
this limit. See an  example of the implementation of the  loop induced Higg boson
vertices in the ``\verb|SM(CKM=1 with hGG/AA)|" model coming with \CalcHEP.
The SM Higgs widths calculated by \CalcHEP~ and Hdecay  
package are
presented in Tab.\ref{tab:Hbranchings} which also present for comparison the respective widths uncertainties 
evaluated in \cite{Dittmaier:2012vm}.

 Here Higgs decays via virtual W/Z are
included in \CalcHEP~ calculation which allows to compare branchings. 
\begin{table}
\begin{center}
\begin{tabular}{|l|l|l|l|l|l|l|}
\hline
          & \multicolumn{3}{|c|}{ Mh=120}& \multicolumn{3}{|c|}{ Mh=150}   \\ 
\hline   
Channel           & CalcHEP & Hdecay &$\pm$\%(th)  & CalcHEP & Hdecay   &$\pm$\%(th)        \\
\hline
$ b\bar{b}$       & 6.55E-01        &  $6.48\cdot10^{-1}$  & 2.8      & 1.59E-01      &$ 1.57\cdot10^{-1}$   & 4.0   \\
$ c\bar{c}$       & 3.33E-02        &  $3.27\cdot10^{-2}$  & 12.2     & 8.05E-03      &$ 7.93\cdot10^{-3}$   & 9.7  \\
$ \tau\bar{\tau}$ & 7.21E-02        &  $7.04\cdot10^{-2}$  & 6.1      & 1.82E-02      &$ 1.79\cdot10^{-2}$   & 3.1    \\
$ZZ$              & 1.48E-02        &  $1.59\cdot10^{-2}$  & 4.8      & 7.76E-02      &$ 8.25\cdot10^{-2}$   & 0.9   \\
$WW$              & 1.41E-01        &  $1.41\cdot10^{-1}$  & 4.8      & 7.04E-01      &$ 6.96\cdot10^{-1}$   & 0.9   \\
$GG$              & 8.16E-02        &  $8.82\cdot10^{-2}$  & 10.6     & 3.17E-02      &$ 3.46\cdot10^{-2}$   & 7.9  \\
$\gamma\gamma$    & 2.30E-03        &  $2.23\cdot10^{-3}$  & 5.4      & 1.39E-03      &$ 1.37\cdot10^{-3}$   & 2.1    \\
\hline
\end{tabular}
\end{center}
\label{tab:Hbranchings}
\caption{SM Higgs boson partical widths from  \CalcHEP~ and Hdecay together with theoretical 
uncertainties  from  \cite{Dittmaier:2012vm}.}
\end{table}
For  $h\to GG$ our result is slightly
smaller  because Hdecay contains NNNLO QCD corrections and  QCD-EW one. For
Higgs mass 120GeV the corrections are 1\% and 5\% respectively. The
last correction is not universal and depends on the model.

The function presented in this section were  included in SLHAplus package 
specially   updated for  \CalcHEP.
  
\subsection{Packages for implementing new models}
\label{modelGeneration}
There can be hundreds or thousands of Feynman rules to implement into
\CalcHEP~for a given model.  This can be a tedious and error prone
process.  For this reason, there have been several packages developed
to take a Lagrangian, derive the Feynman rules and output them to the
format of \CalcHEP.  The first of these was
 {\tt LanHEP}\cite{Semenov:2008jy}.  All of the default models that
 ship with \CalcHEP~(as
 well as many others) were generated using LanHEP.  Recently,
 FeynRules~\cite{Christensen:2008py} and
SARAH~\cite{Staub:2008uz} were created to do a similar job.  Each
package has its strengths.  In this section, we will describe a very
simple extension of the SM and describe its implementation into LanHEP and
FeynRules.

The model we will implement is called the Inert Doublet Model
(IDM)\cite{Barbieri:2006dq,Honorez:2010re}.  In this paper, we will
only describe the new particles and interactions of this model.  We
begin with the SM and add to it a new $SU(2) \times U(1)$ scalar doublet. In unitary
gauge the SM Higgs and the new scalar doublet are given by 
\begin{equation}
\label{IDMdoublets}
 H_1=\left( \begin{array}{c}
  0\\
  \langle v \rangle + h/\sqrt{2}
\end{array}\right)\;\;,\;\;
 H_2=\left( \begin{array}{c}
  \widetilde{H}^+\\
  (\widetilde{X}+i\widetilde{H}_3)/\sqrt{2}
\end{array}
\right)
\end{equation}
where $H_1$ is the SM Higgs doublet and $H_2$ is the new {\it inert} doublet 
which does not couple to quarks and leptons. Unlike the SM scalar doublet it does 
not develop a vacuum expectation value.
$\widetilde{H}^+$, $\widetilde{X}$, and $\widetilde{H}_3$ are the new fields
of the model. The IDM Lagrangian contains only even powers of the doublet $H_2$
\begin{eqnarray}
\label{IDMlagrangian}
   {\cal L}&=&{\cal L}_{SM}+ D^{\mu}H_2^*D_{\mu}H_2 -\mu_2^2 |H_2^2|^2  \\
   \nonumber
   && -\lambda_2 |H_2|^4 -\lambda_3 |H_1|^2|H_2|^2
 -\lambda_4 |H_1^\dagger H_2|^2  - \lambda_5 Re[(H_1^\dagger H_2)^2]
\end{eqnarray}
 Because of the   $H_2 \to -H_2$ symmetry,   the lightest   new particle  is stable. 
In the following example we will use the masses of the new particles as well as 
$ \lambda_2$ and $\lambda_L=(\lambda_3+\lambda_4+\lambda_5)/2$ as independent
parameters.
 The couplings  $\mu$, $\lambda_3$, $\lambda_4$, and $\lambda_5$  
can be expressed in terms of the independent parameters.

\subsubsection{LanHEP} \label{sec:LanHEP}
The  {\tt LanHEP} package was the first package to  automatically
generate 
\CalcHEP~model files from a Lagrangian. It starts from a model definition in terms of particle multiplets
and performs substitutions of physical particle fields from the multiplets.
LanHEP also checks at the symbolic level for the   absence of linear terms  and at the
numerical level for the  absence  of off-diagonal terms in the
quadratic part of the Lagrangian.
Also, LanHEP compares the diagonal quadratic terms with the declared particle
masses.   LanHEP  allows to avoid a large number of mistakes
which could appear in the derivation of Feynman rules by hand. The
package and manual can be found at
\begin{center}
\verb|http://theory.sinp.msu.ru/~semenov/lanhep.html|
\end{center}
The downloaded {\tt lhep}{\it NNN}{\tt .tgz}  file contains  the
source  code as well as a set of 
examples. For our example,  we present part of a LanHEP input file for the
Inert Doublet Model.  We describe only the
new particles and new interactions beyond  the Standard Model.

The LanHEP source file for the IDM should  contain a description of
the new free  parameters
{\footnotesize
\begin{verbatim} 
parameter MHX=111,MH3=222,MHC=333.  % Declaration of new masses  
parameter laL=0.01, la2=0.01.       % Declaration of new couplings 
\end{verbatim}
}
and a declaration of the new  constrained parameters
{\footnotesize
\begin{verbatim}
%mu^2 as a function of masses
parameter mu2=MHX**2-laL*(2*MW/EE*SW)**2.   
% constraints for couplings
parameter la3=2*(MHC**2-mu2)/(2*MW/EE*SW)**2.   
parameter la5=(MHX**2-MH3**2)/(2*MW/EE*SW)**2.
parameter la4=2*laL-la3-la5.                   
\end{verbatim}
}
The new particles of the model would be written as
{\footnotesize
\begin{verbatim}
scalar '~H3'/'~H3':('odd Higgs',pdg 36, mass MH3, width wH3 = auto).
scalar '~H+'/'~H-':('Charged Higgs',pdg 37,mass MHC,width wHC=auto).
scalar '~X'/'~X':('second Higgs',pdg 35,mass MHX,width wHX=auto).
\end{verbatim}
}
We use the component fields defined above to construct the second
scalar doubet as
{\footnotesize
\begin{verbatim}
let h2 = { -i*'~H+',  ('~X'+i*'~H3')/Sqrt2 },
    H2 = {  i*'~H-',  ('~X'-i*'~H3')/Sqrt2 }.
\end{verbatim}
}
Next, we define  covariant derivatives  for the new  doublet where  
{\tt B1} is the SM $U(1)$ gauge field and 
{\tt WW=\{W-,W3,W+\}}  is a SM  $SU(2)$ triplet. {\tt g1} and {\tt g}
are the
corresponding couplings, {\tt taupm} is the generator of the $SU(2)$
group in the
{\tt \{W-,W3,W+\}} basis.  
{\footnotesize
\begin{verbatim}
let Dh2^mu^a = (deriv^mu+i*g1/2*B1^mu)*h2^a +
         i*g/2*taupm^a^b^c*WW^mu^c*h2^b.
let DH2^mu^a = (deriv^mu-i*g1/2*B1^mu)*H2^a
        -i*g/2*taupm^a^b^c*{'W-'^mu,W3^mu,'W+'^mu}^c*H2^b.
\end{verbatim}
}
The terms of the Lagrangian in Eq. (\ref{IDMlagrangian}) can now be
written in LanHEP notation as
{\footnotesize
\begin{verbatim}
lterm DH2*Dh2.                 % Kinematic and other terms.
lterm -mu2*h2*H2.
lterm -la2*(h2*H2)**2.
lterm -la3*(h1*H1)*(h2*H2).
lterm -la4*(h1*H2)*(H1*h2).
lterm -la5/2*(h1*H2)**2 + AddHermConj.
\end{verbatim}
}
where {\tt h1} and
{\tt H1} are the  SM Higgs doublet and its conjugate while
 {\tt h2} and {\tt H2} are the new doublet and its conjugate.

Finally, compilation of the  LanHEP  source file to obtain the
\CalcHEP~model files can be done with the command 
\begin{verbatim}
     lhep  <source file>  -ca -evl 2
\end{verbatim}

\subsubsection{FeynRules}\label{sec:FeynRules}
FeynRules was originally developed to generate MadGraph model files
from a Lagrangian in analogy to LanHEP for \CalcHEP.  However, it was
soon decided that a package that could export to any code was
desirable.  For this reason, one of us (NC) became involved in the
FeynRules project and wrote the \CalcHEP~export functionality for it.
FeynRules has now become a mature package with many of its own
features which support the implementation of new models beyond the
SM.  Further details and a manual can be found at the FeynRules
website:
\begin{center}
\verb|http://feynrules.irmp.ucl.ac.be/|
\end{center}
FeynRules model files use Mathematica notation.
We will now describe the analogous implementation of the IDM using FeynRules.

The model implementation can reuse the SM file (\verb|SM.fr|) which
comes with FeynRules.  To implement the other particles, parameters
and Lagrangians, the user can open a new file in their text editor.
Let's call it \verb|IDM.fr|.  The first thing to define are the
parameters which are enclosed in an array named \verb|M$Parameters| as in
{\footnotesize
\begin{verbatim} 
M$Parameters = {
laL=={	ParameterType -> External,  Value  -> 0.01        },
la2=={	ParameterType -> External,  Value  -> 0.01        },
mu2=={	ParameterType -> Internal,  Value  -> MHX^2-laL*(2*MW/EE*sw)^2,
        Description  -> "mu^2 as a function of masses"    },
la3=={	ParameterType -> Internal,  Value  -> 2*(MHC^2-mu2)/(2*MW/EE*sw)^2,
        Description  -> "constraints for couplings"       },
la5=={	ParameterType -> Internal,   Value -> (MHX^2-MH3^2)/(2*MW/EE*sw)^2},
la4=={	ParameterType -> Internal,   Value -> 2*laL-la3-la5               }
};
\end{verbatim}
}
where the independent parameters are given the property
\verb|ParameterType->External| and a numerical value while the dependent parameters are given
the property \verb|ParameterType->Internal| and an expression for the value.
The new particles of the model would be enclosed in a
\verb|M$ClassesDescription| and written as
{\footnotesize
\begin{verbatim}
M$ClassesDescription = {
S[21] == {
    ClassName        -> X,
    SelfConjugate    -> True,
    Mass             -> {MHX,111},
    Width            -> {wHX,0},
    PDG              -> 35,
    ParticleName     -> "~X",
    FullName         -> "second Higgs"
  },
S[22] == {
    ClassName        -> H3,
    SelfConjugate    -> True,
    Mass             -> {MH3,222},
    Width            -> {wH3,0},
    PDG              -> 36,
    ParticleName     -> "~H3",
    FullName         -> "odd Higgs"
  },
S[23] == {
    ClassName        -> HC,
    SelfConjugate    -> False,
    Mass             -> {MHC,333},
    Width            -> {wHC,0},
    QuantumNumbers   -> {Q -> 1},
    PDG              -> 37,
    ParticleName     -> "~H+",
    AntiParticleName -> "~H-",
    FullName         -> "Charged Higgs"
  },
S[24] == { 
    ClassName      -> h2, 
    Unphysical     -> True, 
    Indices        -> {Index[SU2D]},
    FlavorIndex    -> SU2D,
    SelfConjugate  -> False,
    QuantumNumbers -> {Y -> 1/2},
    Definitions    -> { h2[1] -> -I HC, h2[2] -> (X + I H3)/Sqrt[2]}
}
};
\end{verbatim}
}
where \verb|h2| is defined as the doublet which contains the component
fields.  The antiparticles are automatically defined as are the
covariant derivatives in this simple example.

The terms of the Lagrangian in Eq. (\ref{IDMlagrangian}) can now be
written in FeynRules notation as
{\footnotesize
\begin{verbatim}
LIDM1 = DC[h2bar[ii], mu] DC[h2[ii], mu];
LIDM2 = -mu2^2 h2bar[ii] h2[ii];
LIDM3 = -la2 h2bar[ii] h2[ii] h2bar[jj] h2[jj];
LIDM4 = -la3 Phibar[ii] Phi[ii] h2bar[jj] h2[jj];
LIDM5 = -la4 h2bar[ii] Phi[ii] Phibar[jj] h2[jj];
LIDM6 = -la5/2 h2bar[ii] Phi[ii] h2bar[jj] Phi[jj];
LIDM7 = HC[LIDM6];
LIDM  = LIDM1 + LIDM2 + LIDM3 + LIDM4 + LIDM5 + LIDM6 + LIDM7;
\end{verbatim}
}
where {\tt h1} and
{\tt H1} are represented by {\tt Phi} and {\tt Phibar} while
 {\tt h2} and {\tt H2} are represeneted by {\tt h2} and {\tt h2bar}.

To compile the code and generate \CalcHEP~source code, the user must
first load FeynRules as in
\begin{verbatim}
$FeynRulesPath = "<FR path>";
SetDirectory[$FeynRulesPath];
<< FeynRules`;
\end{verbatim}
where \verb|<FR path>| is the path to the FeynRules package.  Next,
the user must load the model through the following command:
\begin{verbatim}
SetDirectory[<IDM path>];
LoadModel["SM.fr", "IDM.fr"];
\end{verbatim}
where \verb|<IDM path>| is the location of the FeynRules files for the
SM as well as the IDM file described here.  After loading the model,
there are many things that can be done with it in a Mathematica
session such as checking the hermiticity of the Lagrangian,
diagonalization of the quadratic terms, numerical values of the masses
and so on.  Further details can be found in the FeynRules manual.  To
generate \CalcHEP~model files, the user should finally run
\begin{verbatim}
     WriteCHOutput[LSM, LIDM]
\end{verbatim}
which will produce the \CalcHEP~files in the IDM directory.
WriteCHOutput takes several options which are described in the
FeynRules manual.

\subsubsection{SARAH}
SARAH\cite{Staub:2008uz} is designed to work with SUSY models.   Further details can be
found at the SARAH website:
\begin{center}
\verb|http://sarah.hepforge.org/|
\end{center}

\subsection{HEPMDB model repository\label{HEPMDB}}
A convenient location to find and store \CalcHEP~model files is at the 
High Energy Physics Model Database (HEPMDB) \cite{Leshouches2011}  which is located at
\begin{center}
\verb|http://hepmdb.soton.ac.uk/|
\end{center}
HEPMDB is developing quickly and adding new features at a steady rate.
A list of some representative  models currently stored at
HEPMDB along with their respective URL's can be found in
Table~\ref{hepmdb-models}.
\begin{table}[htb]
\begin{tabular}{|l|l|}
\hline
\hline
Model & HEPMDB {\it ID} at \\
      & \verb|http://hepmdb.soton.ac.uk/|{\it ID}\\
\hline
Standard Model (CKM=1) 
&\verb|hepmdb:1211.0043| 
\\ \hline
Standard Model         
& \verb|hepmdb:1211.0042| 
\\ \hline
Littlest Higgs Model with T-parity 
& \verb|hepmdb:0611.0024|
\\ \hline
minimal B-L 
&\verb|hepmdb:0611.0030|
\\ \hline
Minimal B-L with Higgs 1 loop vertices 
&\verb|hepmdb:0611.0031|
\\ \hline
Minimal Zp models 
&\verb|hepmdb:1111.0038|
\\ \hline
MSSM 
&\verb|hepmdb:1211.0028|
\\ \hline
NMSSM (from CalcHEP group)
&\verb|hepmdb:1211.0046|
\\ \hline
NMSSM  with Flavor violation
&\verb|hepmdb:1111.0033|
\\ \hline
NMSSM without Flavor violation
&\verb|hepmdb:1111.0034|
\\ \hline
BLE-SSM (minimal SUSY $U(1)_R \times U(1)_B-L$ &\\
model with inverse seesaw)
&\verb|hepmdb:0611.0075|
\\ \hline
SMSSM (most general, singlet extended MSSM) 
& \verb|hepmdb:0611.0074|
\\ \hline
SUSY inverse seesaw (MSSM with additional &\\
singlet fields to incorporate inverse seesaw)
&\verb|hepmdb:0612.0073|
\\ \hline
B-L-SSM (minimal, supersymmetric B-L model)
&\verb|hepmdb:0612.0072|
\\ \hline
TNMSSM (Triplet extended NMSSM)
&\verb|hepmdb:0612.0072|
\\ \hline
MSSM with bilinear R-Parity violation 
&\verb|hepmdb:1111.0036|
\\ \hline
Model : RPV MSSM (FeynRules)
&\verb|hepmdb:0212.0049|
\\ \hline
RPV MSSM (LanHEP) 
&\verb|hepmdb:0312.0060|
\\ \hline
RPV MSSM (slha) 
&\verb|hepmdb:0512.0068|
\\ \hline
RPV MSSM (softsusy) 
&\verb|hepmdb:0512.0070|
\\ \hline
Model with 1st generation of the Leptoquarks
&\verb|hepmdb:0612.0076|
\\ \hline
Model with 2nd generation of the Leptoquarks
&\verb|hepmdb:0612.0077|
\\ \hline
Model with 3rd generation of the Leptoquarks
&\verb|hepmdb:0612.0078|\\
\hline
\hline
\end{tabular} 
\caption{\label{hepmdb-models} SOme representative models and their  unique links at HEPMDB} 
\end{table}

HEPMDB is not only a convenient repository for CalcHEP models.  It is
much more general and goes beyond CalcHEP,
aiming to:
\begin{enumerate}
\item collect HEP models for a large number of Matrix Element (ME) generators including
{\sf CalcHEP},
{\sf CompHEP}~\cite{Pukhov:1999gg,Boos:2004kh},
{\sf FeynArts}~\cite{Kublbeck:1992mt,Hahn:2000kx},
{\sf
  MadGraph/MadEvent}~\cite{Maltoni:2002qb,Alwall:2011uj,deAquino:2011ub,Degrande:2011ua},
 {\sf AMEGIC ++/COMIX} within {\sf SHERPA}~\cite{Gleisberg:2003xi,Gleisberg:2008ta}
and {\sf WHIZARD}~\cite{Kilian:2007gr}.

\item store the source code for the models which users can download
  or use directly  on HEPMDB to generate HEP models for various ME generators
using the {\sf FeynRules}~\cite{Christensen:2008py} package or the
{\sf LanHEP}~\cite{Semenov:2008jy} package.  The source code consists
of definitions for the particles, parameters and the Lagrangian in
either {\sf FeynRules} or  {\sf LanHEP} format.
\item allow users to upload their models onto the HEPMDB server and
  use the High Performance Computing (HPC) cluster at Southampton
  University (IRIDIS3) to perform evaluations of HEP processes and generate
  events.  The user does this through a web interface on HEPMDB which
  takes care of submitting the jobs and returning the results to the
  user.  IRIDIS3 is currently the largest university owned HPC
  resource in the UK and gives the user of HEPMDB substantial
  computing resources for their calculation.  Thus, HEPMDB provides  a
  web  interface to a large assortment of HEP models, a strong list of 
ME generators and a powerful HPC cluster to use them on.  In this way,
the user can bypass the difficulties involved in installing and
setting up the
various software and, instead, focus on the physics.
\item 
 cross check and validate new model implementations between different
 ME generators and different gauges. We should note
 that similar functionality is also provided by the FeynRules web
 validation \cite{Leshouches2011}.  However, although the FeynRules
 web validation is more highly automated, it is focused on FeynRules
 models whereas the HEPMDB validation is designed to be more general
 and allow models implemented with any package to be validated.
Also, it should be noted that the FeynRules web validation is only
designed to
perform validations.  It does not have any capability to perform
general physical processes or generate events.


\item systematically collect the predictions and detailed features of
  the models included in the database.  This will involve the
  development of a comprehensive database of signatures and the
  development of a format for the presentation of these features.
The format for the  signatures will be consistent with the format used
by the experiments and    
HEPMDB wll perform a comparison of the resulting predictions of
  the models with the experimental data.  
Further details can be found in the
Les Houches report \cite{Leshouches2011} under the title 
``Les Houches Recommendations for the Presentation of LHC Results" activity.

\item 
trace the history of the modifications of each model and make the full
history available to the user. 
This feature will enable reproducibility for the results obtained from
the models stored on HEPMDB.  

\end{enumerate}

\section{\label{sec:getMEcode}Using \CalcHEP~ as a matrix elements 
generator  for other packages.}

In the previous sections, we have discussed the use of the
\CalcHEP~generated code within \CalcHEP.  There are times, however,
when a user would like to use the optimized squared matrix element code with an
external package or with \hisher~own code.  In the current version,
a code has been included to allow new processes to be generated and
linked dynamically.  This  code was first written for the
micrOMEGAs\cite{Belanger:2006is, Belanger:2010gh} package.  In this section, we
describe how to generate and use these dynamic libraries.

To generate
a squared matrix element for a given process dynamically, 
the user should write a main C
program.  An example of such  program  can be found at
\verb|$CALCHEP/utile/main_22.c|.   We have also included a shell
script to compile the main program collecting all the necessary
libraries.  It can be used as in the following example:
\begin{center}
\verb|$CALCHEP/bin/make_main [-o<exe_name>]  <C source codes and libraries>  |
\end{center}
If compilation was successful, the excutable  will be created and can be
run.   In  this section, we describe the basic elements that are
involved in writing a main program which can be extended for more
complicated cases.

\subsection{\label{sec: set up}Set up}
At the beginning of the main file, the user should include all the
required headers including the following \CalcHEP~headers:
\begin{verbatim}
#include"num_in.h"
#include"num_out.h"
#include"VandP.h"
#include"dynamic_cs.h"
#include"rootDir.h" 
\end{verbatim}

\subsection{Model choice}
The first thing the user must do is to choose the \CalcHEP~model
\heshe~will work with.  This is done with the \verb|setModel| which
has the following prototype
\begin{verbatim}
int  setModel(char* modelDir, int n)
\end{verbatim}
where \verb|modelDir| is a string containing the path to the models
directory and \verb|n| is the model number for the model.  For
example, if the main program was being written in the \verb|WORK|
directory, this function call might look like
\verb|setModel("models",1)|.  
The \verb|setModel| command generates the \verb|aux| subdirectory which is
organised as a \CalcHEP~working directory with the subdirectories 
\verb|models, results, tmp| and \verb|so_generated| where the process libraries will
be stored.  It is very important to note that the model files in the
\verb|aux| directory will be copies of the ones from the
\verb|WORK/models| directory.  After it has been created, any changes
to the model in \verb|WORK/models| will not affect those in \verb|aux|
and vice versa.  
The \verb|setModel| function will also create the library
\verb|VandP.so| in the \verb|aux| directory which contains a list of
particles from the model as well as the compiled parameters of the
model.
If \verb|setModel| is successful, it returns 0.

It is possible to use multiple models in the same main function, but
in this case, the \verb|aux| directory will be cleaned and any
libraries will be lost and have to be recreated.

\subsection{Model parameters}
\label{setting_parameters}

Typically, the user would like to change the numerical value of the
parameters, perhaps in a scan over parameters.  There are three
functions which can be used for this purpose.  The prototypes for the
first two are:
\begin{verbatim}
int assignVal(char* name, double val)
void assignValW(char* name, doube val)
\end{verbatim}
where \verb|name| is a string containing the independent parameter
name and \verb|val| is the numerical value for that parameter.
The only difference between these two functions is what happens when
an error occurs.  In the first case, a nonzero integer is returned
while in the second case an error message is printed to
stderr.

Additionally, the new parameter values can be read from a text file
with the following function
\begin{verbatim}
int readVar(char* fileName)
\end{verbatim}
where \verb|filename| is the file that contains the new values.  The
file must contain two columns separated by whitespace.  The first
column must contain the name of the parameter while the second column
must have the numerical value for that parameter.  For example
\begin{verbatim}
Mh     125
MA     500
...
\end{verbatim}
where \verb|...| represent further lines with other parameter values.
\verb|readVar| returns 0 upon success.   Otherwise, it returns a negative value when the file
cannot be opened or a positive value corresponding with the line
number in the file which cannot be read.

After the independent parameters are assigned, the user must call the function
\begin{verbatim}
int  calcMainFunc(void)
\end{verbatim}
which calculates and updates all the {\it public} (see 
Section \ref{constraints} for a definition of public) dependent parameters.  This
function returns 0 when successful.  If an error occurs, it returns a
positive integer \verb|err| which corresponds with the dependent
parameter (\verb|varNames[err]|) where the error occurred.

It is often useful to view the updated values of the parameters.
For this reason, we include the following functions which work for any
independent parameter or any {\it public} dependent parameter
\begin{verbatim}
int  findVal(char* name, double* val)
double  findValW(char* name)
\end{verbatim}
where \verb|name| is a string with the name of the parameter.  The
first function, \verb|findVal|, sets \verb|*val| equal to the
numerical value of the paremeter and returns a nonzero value if it
cannot find the parameter.  The
second function, \verb|findValW| simply returns the value of the
parameter.  It prints an error message to stderr if it cannot locate
the parameter.


\subsection{\label{sec: model particles}Model Particles}

The properties of the particles (see Section \ref{sec: symbolic}) can be obtained with the functions in
this section.  The particle name can be obtained from the PDG code by
using the function
\begin{verbatim}
char* pdg2name(int nPDG)
\end{verbatim}
where \verb|nPDG| is the PDG code.  This function returns a string
with the particle name upon success and \verb|NULL| otherwise.
On the other hand, the PDG code can be obtained from the name of the
particle with the function
\begin{verbatim}
int pNum(char * name)
\end{verbatim}
which takes the string \verb|name| and returns the PDG code.
However, if the particle name cannot be found, it returns 0.

The quantum numbers of the particle can be obtained with the function
\begin{verbatim}
int  qNumbers(char* pName, int* spin2, int* charge3, int* cdim)
\end{verbatim}
where the string \verb|pName| should be the particle's name.  The
variable \verb|*spin2| will be filled with twice the spin of the
particle, \verb|*charge3| will be filled with three times the electric
charge of the particle, and \verb|*cdim| will be filled with the color
representation of the particle (either \verb|1,3,-3| or \verb|8|).
This function will return the PDG code of the particle upon success
and 0 otherwise.

The numerical value of the mass of a particle is returned by the function
\begin{verbatim}
double *  pMass(char* pName)
\end{verbatim}
where the string \verb|pName| is the name of the particle.

\subsection{Direct access to model parameters and particles description}
At times, it will be necessary to access the parameters directly.
They are stored in the following variables
\begin{verbatim}
int nModelVars;
int nModelFunc;  
char **varNames; // contains nModelVars+nModelFunc+1 elements; varNames[0] is not used.
REAL *varValues; // contains nModelVars+nModelFunc+1 elements; varValues[0] is not used.
\end{verbatim}
where \verb|nModelVars| is the number of independent parameters in the
model, \verb|nModelFunc| is the number of {\it public} dependent
parameters, \verb|varNames| is an array of strings containing the
parameter names, and \verb|varValues| is an array containing the
parameter values.    It is important to note that both \verb|varNames|
and \verb|varValues| have \verb|nModelVars+nModelFunc+1| elements.
Both \verb|varNames[0]| and \verb|varValues[0]| are not used.
The Type \verb|REAL| is defined in the file
\verb|$CALCHEP/include/nType.h|. 
By default
\verb|REAL| corresponds with \verb|double|.

If the particle properties have to accessed directly, the variables
for this are
\begin{verbatim}
int nModelParticles;
ModelPrtclsStr *ModelPrtcls;
\end{verbatim}
where \verb|nModelParticles| is the number of particles in the model
and the type \verb|ModelPrtclsStr| is
defined in \verb|$CALCHEP/include/VandP.h|

\subsection{Decay widths and branching fractions}
\label{widths}
The calculation of particle widths, decay channels  and branching fractions
can be done with the function
\begin{verbatim}
double  pWidth(char* pName, txtList*branchings)
\end{verbatim}
where \verb|pName| is the name of the particle.  This function returns
the width of the particle and fills the variable
\verb|branchings| with the details of the decay channels.  The type \verb|txtList| is defined in the file 
\verb|$CALCHEP/c_sources/dynamic_me/include/dynamic_cs.h|.   
The use of \verb|branchings| will be described below.  

If the widths of the particles were read from a SLHA file before the
calling \verb|pWidth|, then the width from the SLHA file will
immediately be returned.  On the other hand, if the widths were not
read from an SLHA file, \CalcHEP~will calculate them in the following way.
\CalcHEP~will first calculate the partial
widths from all kinematically open $1\to2$
decays channels.  If the resulting width is zero at this point, it
calculates the partial widths from all kinematically open $1\to3$
decay channels.  If the width is still zero, it tries the $1\to4$
decay channels.  If the width is still zero, it returns zero for the
particle's width.  Otherwise, it returns the calculated width.
Variable \verb|VVdecay| controles account of decay channels with vitrial
W/Z, Section \ref{sec: symbolic}. By default \verb|VVdecay=1| and these 
channels are contribute to particle widths.  

Once the decay widths and branching ratios are calculated, they can be
printed to file using the function 
\begin{verbatim}
void printTxtList(txtList branchings, FILE* f)
\end{verbatim}
where \verb|branchings| is the same variable used in the function
\verb|pWidth| and \verb|f| is the file handle which has already been
opened for writing.

The branching ratio for a specific final state can also be obtained
using the function
\begin{verbatim}
double  findBr(txtList branchings , char* pattern)
\end{verbatim}
where \verb|branchings| is the same variable used in the function
\verb|pWidth| and the string \verb|pattern| should be a comma
separated list of the (anti)particle names in the final state.
The order is not important.  
\verb|findBr| returns the branching ratio for the
requested final state.

An alternate function that calculates the width and branching ratios is
\begin{verbatim}
int  slhaDecayPrint(char* pname, FILE* f)
\end{verbatim}
where \verb|pname| is the name of the particle and \verb|f| is the
open file handle where the width and full list of branchings will be
written in SLHA format.  
The return value is the PDG code of the particle if successful and
zero otherwise.

\subsection{Processes} 

\CalcHEP~contains facilities to dynamically generate and link the code
for any collision or decay process.  This is done with the function
\begin{verbatim}
numout* getMEcode(int twidth, int UG, char* Process, char* excludeVirtual, 
                  char* excludeOut, char* libName)
\end{verbatim}
where\\
{\tt twidth} is a flag which determines whether the t-channel
propagators contain the particle width;\\
{\tt UG} is a flag which forces Unitary Gauge calculations whatever  gage
was  initially specified in the model;\\
{\tt Process} is a string which contains the desired process;\\
{\tt excludeVirtual} is a string containing a comma separated list of
particles to remove from the internal lines of diagrams;\\
{\tt excludeOut} is a string containing a comma separated list of
particles to remove from the final state particles.  This is useful
when the process string contains ``2*x'';\\
{\tt libName} is a string containing the name of the library where
the compiled code will be stored.  This name should not contain the
``.so'' suffix.

If the library already exists, it will be dynamically linked and a
pointer to the code of type \verb|numout| will be returned.
Note that if the library exists, it will not be checked whether it
contains the same process.  The user must take care with the names of
the libraries and ensure they are correctly used.
If the library cannot be found, it will attempt to dynamically generate and compile the
code.  If successful, it will store it in the file
\verb|aux/so-generated/libName.so|
where \verb|libName| is the string given to \verb|getMEcode|.  It will
then dynamically link it and return the pointer to the code.  
Since the string \verb|libName| is used for the name of the file, it
cannot contain any characters not allowed in file names.  For example,
the special characters \verb|+, -, *,| or \verb|/| are not allowed.
If the code cannot be generated and compiled successfully, \verb|NULL|
is returned.  This could occur, for example, if the process is not
allowed in the model.
The structure of the type \verb|numout| can be found in \\
\verb|    $CALCHEP/c_sources/dynamic_me/include/dynamic_cs.h|.\\ 
We will not explain this structure but give some
examples of it usage. 

\CalcHEP~  has a {\it light} version of \verb|getMEcode| \\
\verb|   newProcess(char* Process) |\\
which corresponds to \verb|getMEcode|  with parameters 
\verb|twidth=UG=0|, \verb|excludeVirtual=excludeOut=NULL| and
\verb|libName| defined automatically according to the \verb|Process|
parameter. 

At this point, the parameters stored in dynamically generated process
code are not related to the parameter values described above in
Section \ref{setting_parameters}.  Before we use the dynamically
linked code, it is very important to export numerical values of parameters.  This can be done
by  call 
\begin{verbatim} 
   passParameters(cc);
\end{verbatim}

Further information about the processes in the dynamically linked
library {\tt cc} can be obtained with the functions
\begin{verbatim}
int procInfo1(numout* cc, int* nproc, int* nin, int* nout)
int procInfo2(numout* cc, int nsub, char** pName, REAL* Masses)
\end{verbatim}
Both of these functions return zero if
successful.  In the first function, \verb|*nproc| is filled with the
number of subprocesses in the library, \verb|*nin| is filled with the
number of ingoing particles, and \verb|*nout| is filled with the number of
outgoing particles.  In the second function, $1\le${\tt nsub}$\le${\tt
nproc}  specifies which subprocess the user would like information about, 
\verb|pName| is an array of strings which is filled with the names of the ingoing
and outgoing particles, and \verb|Masses| is an array containing the
masses of the ingoing and outgoing particles.  Both \verb|pNames| and
\verb|Masses| are arrays of size \verb|nin+nout|.  The ingoing
particles are listed first with the outgoing particles following.

\subsection{Matrix elements}
After all the parameters are set and the code is generated and linked,
the matrix element can be obtained by 
\begin{verbatim}
    cc->interface->sqme(int nsub, double GG, REAL* pvect, int* err_code)
\end{verbatim}

  To do this, the user has to first fill the momentum array for the phase space 
point of interest and define strong coupling $GG=\sqrt{4\pi\alpha_{QCD}}$
corresponding to momenta. \verb|pvect| should be an array of $4\cdot(nin+nout)$  
elements which contains subsequently  4-momenta of particles:\\ 
\verb| REAL *p1=pvec, *p2=pvec+4, *p3=pvec+8, *p4=pvec+12 ...|\\
Each momenta contains the energy first followed by the 3
dimensional momentum.  So, for example,\\
$(E_1,p_{1x},p_{1y},p_{1z})=$(\verb|p1[0],p1[1],p1[2],p1[3]|).\\
Here \verb|nsub| determines which subprocess to use and
\verb|*err_code|
is an integer which will contain the error code if any or zero otherwise.
The return value of \verb|sqme| function   is the squared 
matrix element at this phase space point. 
It should be noted that the polarizations of ingoing particles have
already been summed over and the polarizations of outgoing particles
have already been averaged in the result.

Compilation of \verb|main_22.c| should create an  executable. By befault its
name is \verb|a.out|. It
has to be  launched in user's  \verb|$WORK| because it compiles matrix    
elements for models defined by  \verb|models/*2.mdl| files. It  calculates
Fermi constant, width and decay branchings of W+  in SLHA format, compiles  \verb|e,E->m,M|
matrix element and  avaluates  it at  Pcm=100GeV. The screen output  should be 
{\footnotesize
\begin{verbatim}
GF=1.165943E-05
DECAY 24  2.011200E+00  # W+
 1.111241E-01   2   -11 12  #  E,ne 
 1.111238E-01   2   -13 14  #  M,nm 
 1.110418E-01   2   -15 16  #  L,nl 
 3.333724E-01   2   -1 2  #  D,u 
 3.333379E-01   2   -3 4  #  S,c 

sigmaTot(Pcm=1.000000E+02)= 2.955426E+00
\end{verbatim}
}
\subsection{ Matrix elemnts for Root package} 
In this session we explain how dynamicaly defined matrix elements can be
used in Root package. Root has an option to load  user shared library and
supports  interface with C++ classes prepared in special manner. 
A part of function presented above was combined in C++ class 
{\footnotesize
\begin{verbatim}
 class ch_class:public TObject  {
  public:
    txtList L;
    numout*cc;
    int Nin,Nout,Nproc;
    
    ch_class(void);
    int    SetModel(char * modelDisp , int nModel);
    int    AssignValW(char * name, double val);
    int    CalcMainFunc(void);   
    double FindValW(char * name);
    double PMass(char * pName);  
    int    PNum(char*pName);
    char*  Pdg2name(int pdg);
    int    QNumbers(char *pname, int *spin2, int * charge3, int * cdim);
    double PWidth(char * pName);   
    void   PrintTxtList(FILE*f);
    double FindBr(char*pattern);
    int    SlhaDecayPrint(char*pname,FILE*f);    
    int    GetMEcode(int twidth,int UG,char*Process,char*excludeVirtual,char*excludeOut,char*lib);
    int    NewProcess(char*Process);
    int    PassParameters(void);
    char * ProcInfor2(int nsub, char** pName   double*pmass);    
    double Sqme(int nsub, double GG, double *pvect,int*err);
    ClassDef(ch_class, 1)
 };
\end{verbatim}
}

The functions included in this class are close to ones presented above.  To
avoid names conflict we start C++ functions from capital letters. The  C++
functions are slightly different form the original C ones.  We have included
\verb|txtList L| variable in the class and \verb|PWidth| directly assigns
value to it.  From the other side  {\tt PrintTxtList} and { \tt FindBr} use
this parameter implicitly. The same about  variable \verb|numout*cc| which
is filled by  \verb|GetMEcode| or  \verb|NewProcess|  and is used by 
{\tt  PassParameters, ProcInfo2, Sqme } implicitly. {\tt GetMEcode} and   {\tt NewProcess} fill {\tt
Nin,Nout,Nproc} parameters. {\tt CalcMainFunc} calls {\tt PassParameters} if
{\tt cc} is initiated.  For C++ case we assume that \CalcHEP~ uses type {\it
double} for numerical calculations. 

In order to use \CalcHEP~ functions  under Root one has to compile
shared library \verb|ch_root.so|. It can be done by {\it make} command
launched from \verb|$CALCHEP/c_source/Root| directory. Before it one has to
define  destination of Root package via  {\it ROOT} variable. In result of
compilation library \verb|ch_root.so| has to appear in \verb|$CALCHEP/lib|
directory.  Directory {\it utile} contains file \verb|main_22.root| which
actually  is a translation  of example presented above in C++ language.
It is assumed that user passes it  to  {\it Root } interpreter by the command 
\begin{verbatim}
    root < ../utile/main_22.root     
\end{verbatim}
launched from \verb|$CALCHEP/work| directory. The same output as in the C program case is
expected. 

\section{\label{sec-benchmarks} 
Batch file examples with results}

In this section we present some 
examples of calculations using the batch interface (see Section
\ref{sec:batch interface}).  We give the full details of the input and the output so
users can test their installation.  Users can also use these examples
as templates for their own calculations.
We  assume in this section that the user has already installed 
\CalcHEP~and has already created a \verb|WORK| directory.

\subsection{$e^+e^-\to Z h$}
Our first example is for the process $e^+e^-\to Z h$.  The batch file
for this example comes with \CalcHEP~and can be found at 
\verb|$CALCHEP/utile/batch_file_1|.
It has the following form:
{\footnotesize
\\
\\
\begin{spacing}{0.1}
\begin{verbatim}
Model:         Standard Model
Model changed: False
Gauge:         Feynman
Process:       e,E->Z,h

pdf1:      ISR & Beamstrahlung
pdf2:      ISR & Beamstrahlung
Bunch x+y sizes (nm)	  : 560
Bunch length (mm)	  : 0.4
Number of particles	  : 2E+10

p1:        500
p2:        500

Parameter: Mh=125

Dist parameter:    E1
Dist min:	   30
Dist max:	   499
Dist n bins:	   100
Dist title:	   e,E->Z,h
Dist x-title:	   E1 (GeV)

Number of events (per run step): 10000
Filename:                        ee_zh
Max number of cpus:          2

nSess_1:   5
nCalls_1:  100000
nSess_2:   5
nCalls_2:  100000
\end{verbatim}
\end{spacing}
}

\begin{figure}[htb]
\includegraphics[width=0.55\linewidth]{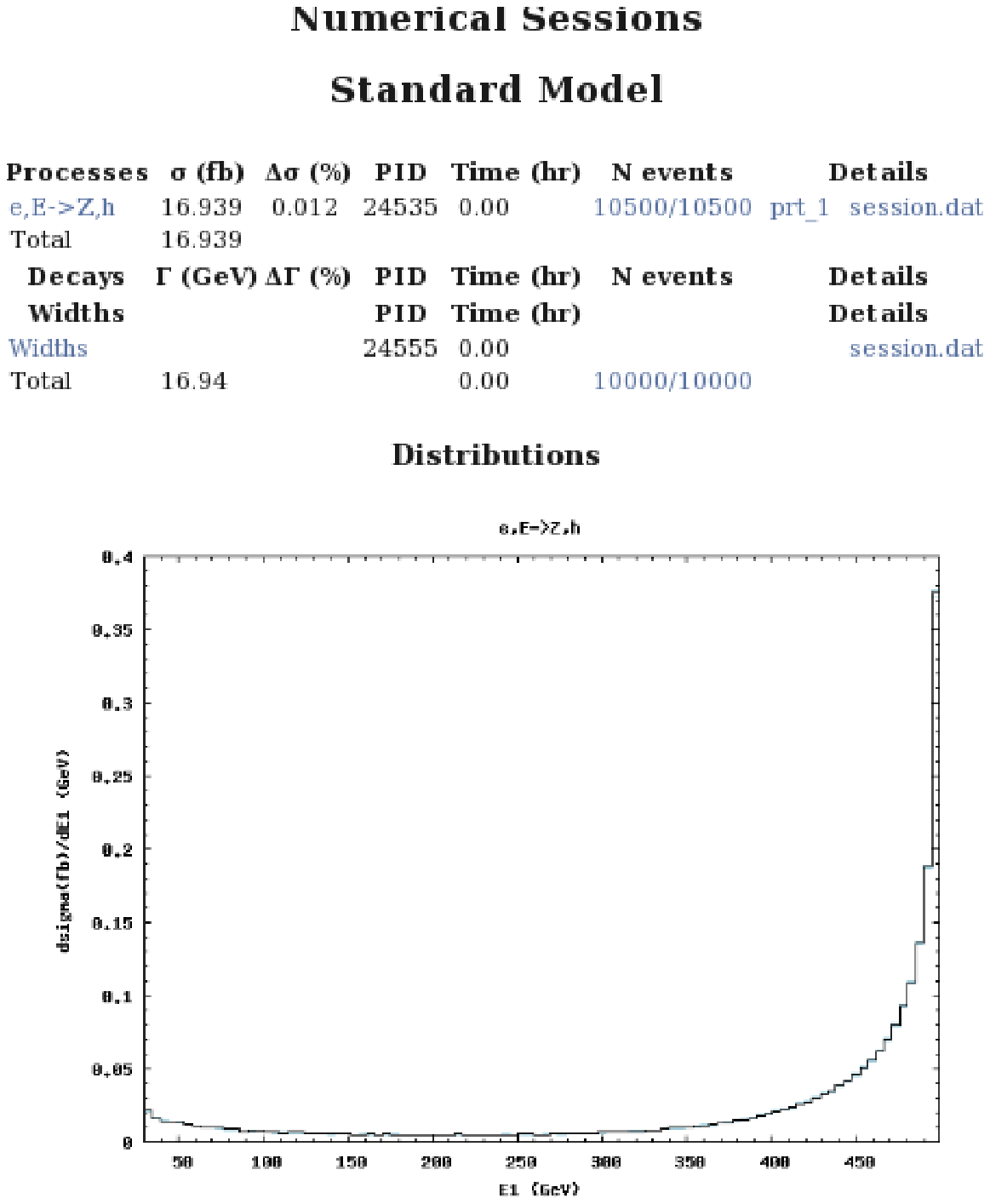}
\caption{\label{batch1} Results of the {\tt batch\_file\_1}
evaluation.}
\end{figure}

This batch example can be run with the following command (from the
\verb|WORK| directory)
\begin{center}
\verb|./calchep_batch batch_file_1|
\end{center}
It will calculate the cross section of the process 
$e^+e^- \to Z h$ at $\sqrt{s}=1$~TeV
taking into account ISR \& Beamstrahlung
effects using the beam geometry specified in the batch file.
This batch will also generate 10K events and write them 
in the file \verb|ee_zh-single.lhe|
located in the \verb|$WORK/Events/| directory in LHEF format.
The total cross section of this process is 16.9 fb.  The result of
this batch session can be observed in a web browser as in
Fig.~\ref{batch1}.

\newpage
\subsection{$e^+e^- \to h\mu^+\mu^- \to \mu^+\mu^- b\bar{b}$  with only $e^+e^- \to Zh$ diagram}
In this example, we will evaluate the process
 $e^+e^- \to h\mu^+\mu^- \to \mu^+\mu^- b\bar{b}$ and remove the
 diagrams that do not correspond to $e^+e^- \to Zh$.  Our batch file
 can be found at \verb|$CALCHEP/utile/batch_file_2|
and has the following form:
\skip 0.5cm
{\footnotesize
\begin{spacing}{0.1}
\begin{verbatim}
Model:         Standard Model
Model changed: False
Gauge:         Feynman
Process:       e,E->m,M,h
Decay:         h-> b,B
Remove:        A,e,m

pdf1:      ISR & Beamstrahlung
pdf2:      ISR & Beamstrahlung
Bunch x+y sizes (nm)	  : 560
Bunch length (mm)	  : 0.4
Number of particles	  : 2E+10

p1:        500
p2:        500

Parameter: Mh=125

Regularization momentum: 34
Regularization mass:     MZ
Regularization width:    wZ
Regularization power:    2

Dist parameter:    M(m,M)
Dist min:	   0
Dist max:	   120
Dist n bins:	   100
Dist title:	   e,E->m,M,h
Dist x-title:	   M(m,M) (GeV)

Number of events (per run step): 10000
Filename:                        ee_mmh_mmbb

nSess_1:   5
nCalls_1:  100000
nSess_2:   5
nCalls_2:  100000
\end{verbatim}
\skip 0.5cm
\end{spacing}
}

One should note three new elements
in this batch as compared to the previous one:\\
1. {\tt Decay:} this statement generates  Higgs boson decay 
which is automatically connected to the production mode;
\\
2. {\tt Remove:} this statement specifies that diagrams with internal
photons, electrons or muons should be removed (that is, only the
$e^+e^- \to Zh$ process is kept).
\\
3. {\tt Regularization:} this statement is used to perform a more efficient integration of the
Z-boson resonance peak.
\\

\begin{figure}[htb]
\includegraphics[width=0.6\linewidth]{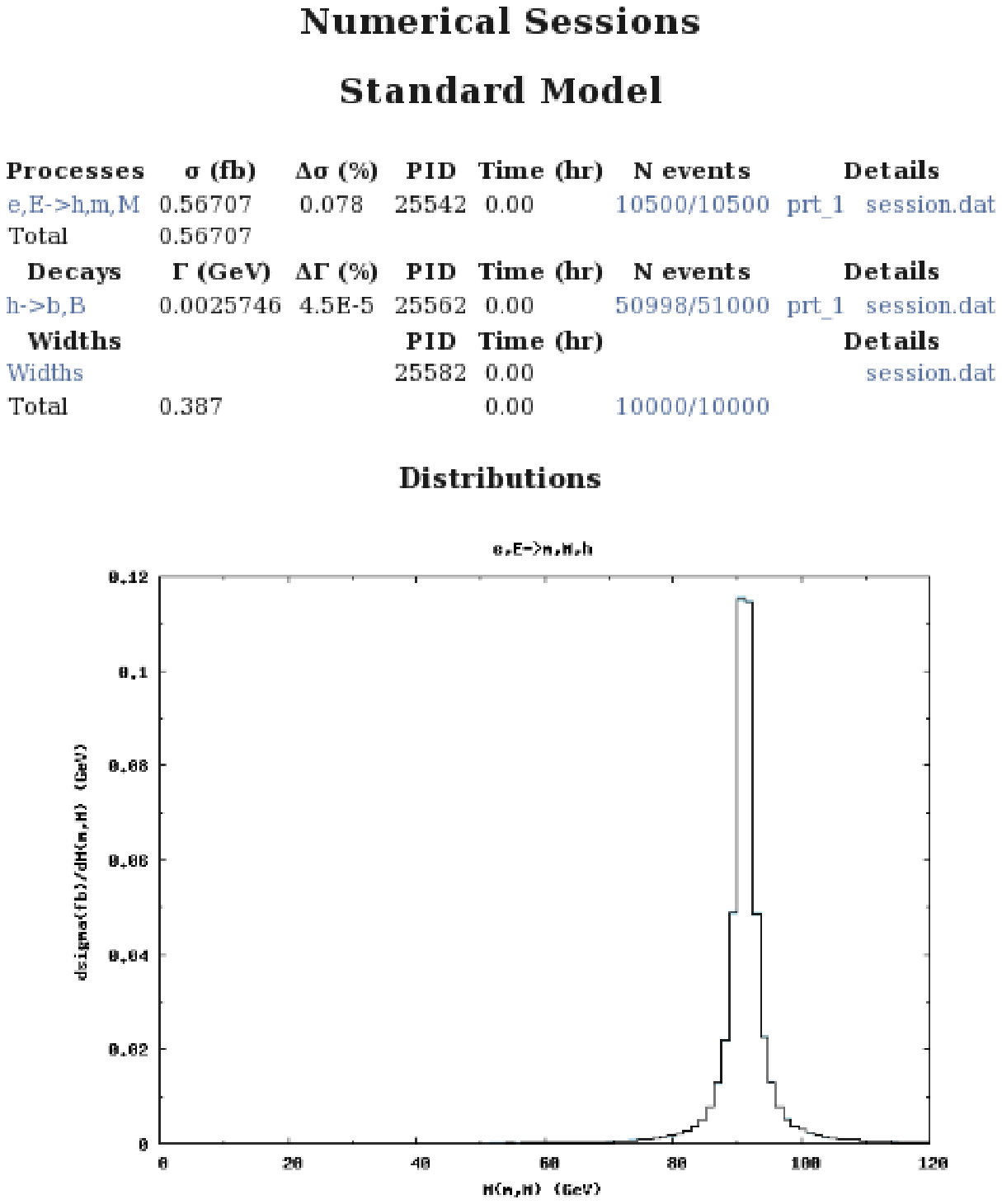}
\caption{\label{batch2} Results of the {\tt batch\_file\_2}
evaluation.}
\end{figure}

By running
\begin{center}
\verb|./calchep_batch batch_file_2|
\end{center}
the batch interface  will calculate the cross section,
generate 10K events with a $\mu^+\mu^- b\bar{b}$ final state
written in the \verb|ee_mmh_mmbb-single.lhe| file.
The total cross section of this process is 0.517 fb.
Again, the user can check the results in \hisher~web browser which
should agree with
Fig.~\ref{batch2}.

\newpage
\subsection{$pp\to Wb\bar{b} \to \ell \nu b \bar{b}$}
In this example, we consider the process
 $pp\to Wb\bar{b} \to \ell \nu b \bar{b}$.  
The batch file can be found at \verb|$CALCHEP/utile/batch_file_3|
and has the following form:
\skip 0.5cm
{\footnotesize
\begin{spacing}{0.1}
\begin{verbatim}
Model:         Standard Model(CKM=1)
Model changed: False
Gauge:         Feynman

Process:   p,p->W,b,B
Decay:     W->le,n
Composite: p=u,U,d,D,s,S,c,C,b,B,G
Composite: W=W+,W-
Composite: le=e,E,m,M
Composite: n=ne,Ne,nm,Nm
Composite: jet=u,U,d,D,s,S,c,C,b,B,G

pdf1:      cteq6l (proton)
pdf2:      cteq6l (proton)
p1:        4000
p2:        4000

Run parameter: Mh
Run begin:     120
Run step size: 5
Run n steps:   3

alpha Q :            M45

Cut parameter:    M(b,B)
Cut invert:       False
Cut min:          100
Cut max:     

Cut parameter:    J(jet,jet)
Cut invert:       False
Cut min:          0.5
Cut max:            

Cut parameter:    T(jet)
Cut invert:       False
Cut min:          20
Cut max:            

Kinematics :      12 -> 3, 45
Kinematics :      45 -> 4 , 5

Regularization momentum: 45
Regularization mass:     Mh
Regularization width:    wh
Regularization power:    2


Dist parameter:    M(b,B)
Dist min:          100
Dist max:          200
Dist n bins:       100
Dist title:        p,p->W,b,B
Dist x-title:      M(b,B) (GeV)

Dist parameter:    M(W,jet)
Dist min:	   100
Dist max:	   200
Dist n bins:	   100
Dist title:	   p,p->W,b,B
Dist x-title:	   M(W,jet) (GeV)

Number of events (per run step): 10000
Filename:                        pp_Wbb_enbb

nSess_1:   5
nCalls_1:  100000
nSess_2:   5
nCalls_2:  100000
\end{verbatim}
\skip 0.5cm
\end{spacing}
}

\begin{figure}[h]
\includegraphics[width=0.6\linewidth]{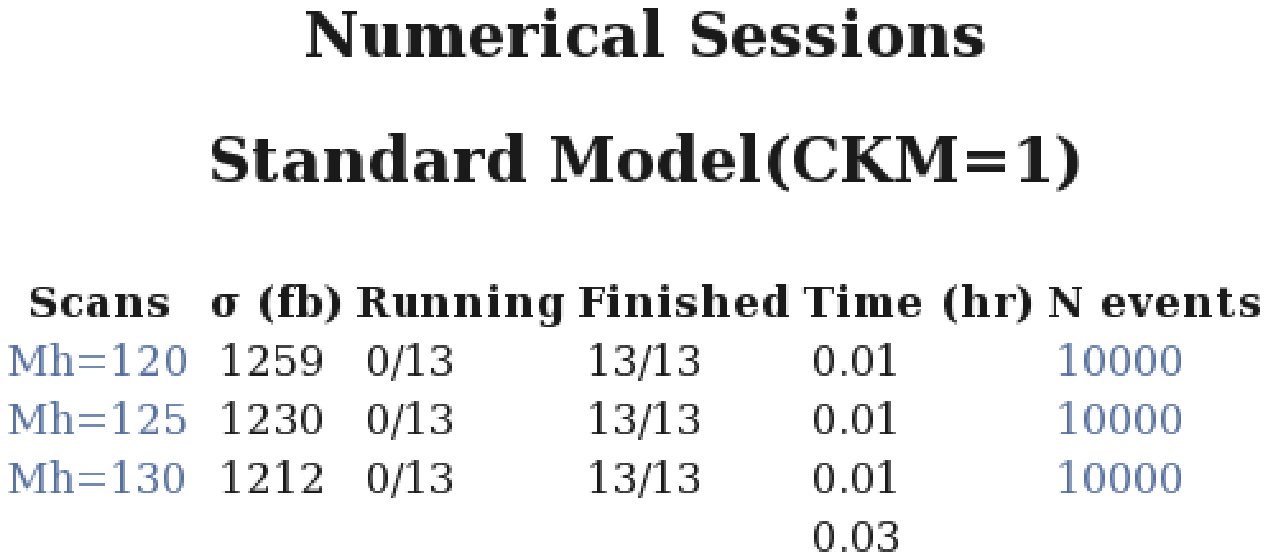}
\caption{\label{batch3} Results of the {\tt batch\_file\_3}
evaluation.}
\end{figure}

A new feature of this batch file is that it performs a scan of  $pp\to Wb\bar{b} \to \ell \nu b \bar{b}$
as a function of the Higgs boson mass
using the  \verb|Run parameter| statement.
The result of this scan can be seen on a web browser as shown in Fig.~\ref{batch3}.

\begin{figure}[htb]          
\includegraphics[width=0.6\linewidth]{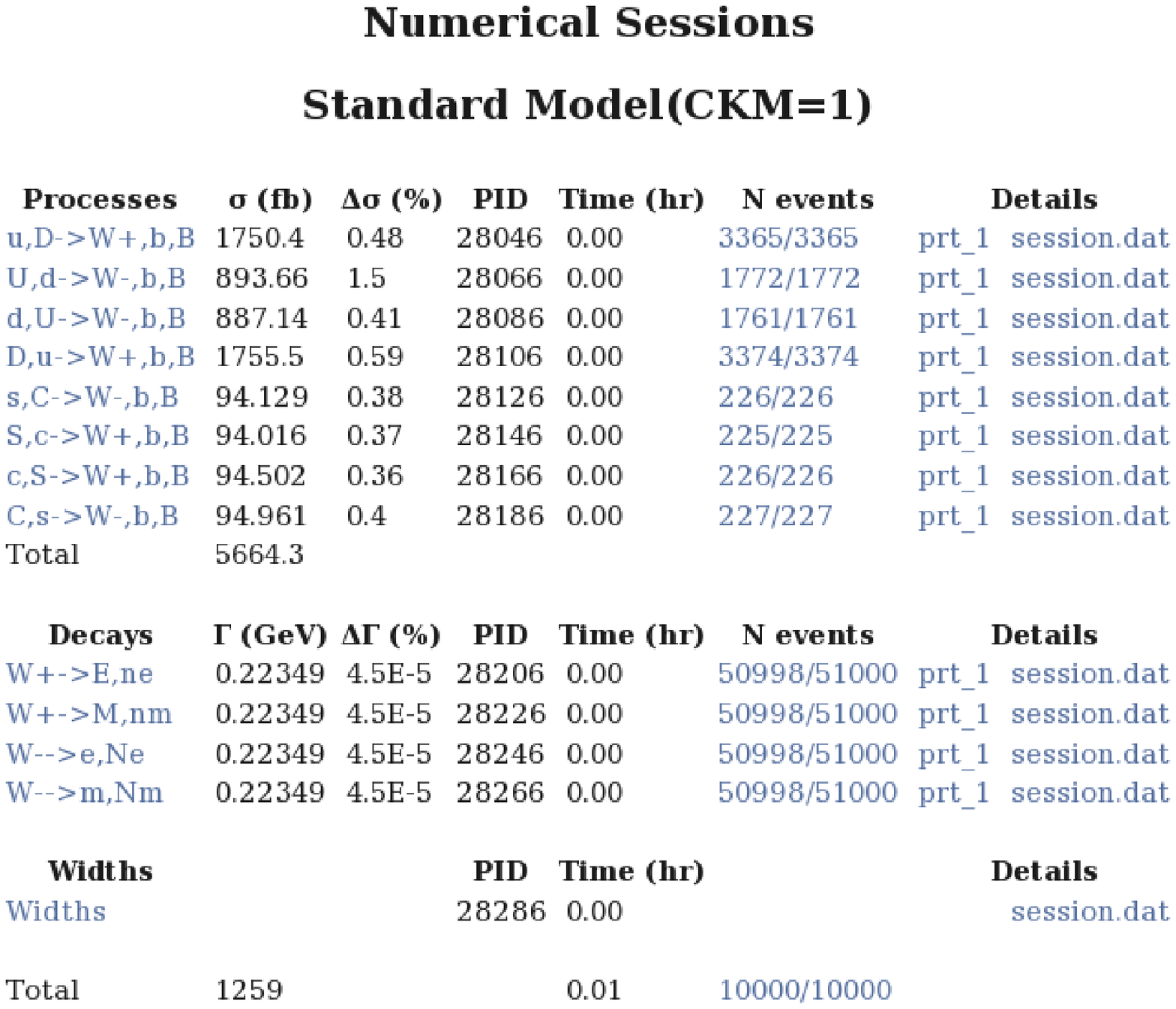}
\caption{Details of the numerical session for $pp\to Wb\bar{b} \to \ell \nu b \bar{b}$
for $M_h=120$~GeV for the  {\tt batch\_file\_3} file evaluation}
\label{batch-num}
\end{figure}

Further details on the individual numerical sessions can be checked by clicking
on a
particular value of the running parameter.  For example, clicking on
\verb|Mh=120| in the web browser will lead to the html page shown in 
Fig.\ref{batch-num}.
This page also presents the requested distributions.

We would like to note that all 
the results shown in the html browser are also recorded in the pure text files
located in the \verb|$WORK/html| directory.
For example, the results for the numerical session are recorded in the file
\verb|$WORK/html/numerical.txt| as well as in the files in the \verb|$WORK/html/runs|
directory which contain further details.
For example, after a successful run of the file  \verb|batch_file_3| given in the example above,
the user should get the file
 \verb|$WORK/html/numerical.txt|  with the following lines:
\begin{verbatim}
	CalcHEP Numerical Details

	Done!

Scans               sigma (fb)     Running   Finished   Time (hr)   N events
Mh120               1.2610e+03       0/13     13/13     0.01        10000
Mh125               1.2510e+03       0/13     13/13     0.01        10000
Mh130               1.2440e+03       0/13     13/13     0.01        10000

\end{verbatim}

\section{\label{sec:conclusions}Conclusions and outlook}

 The new version of \CalcHEP, version~\chvers~presented here is ready 
to explore the Standard Model and  BSM models by theorists, phenomenologists and experimentalists.
\CalcHEP~can be used:
\\
1) via an interactive GUI interface which allows to understand the anatomy of the process under
study including details of the interference;
\\
2) via a batch interface which highly automatizes and parallelizes the evaluation
of numerous production and decay processes as well as conencting them together at the stage of event generaton;
\\
3) through the High Energy Physics Model Database (HEPMDB) using the resources
of a  powerful HPC cluster
which allowes  to perform an exaustive scan of the model parameter space
and generate numerous LHE files in one run.

We should also stress that there are many BSM models available for \CalcHEP 
at the HEPMDB site and it is easy to implement new models by means of the LanHEP and FeynRules packages.

The next development of \CalcHEP~will include  \\
1) an implementation of an automatic regularisation procedure.
\\
2) an implemention of projections on polarization 
states of the massive fermions and  vector bosons 
\\
3) taking into account spin correlatons
for the processes when connecting production and decays.
\\
4) an implementation of the jet matching procedure which will allow to properly combine 
   subprocess with multi-jet final states.
\\
5) an implementation of the helicity amplitude method which allows to evaluate 
processes with larger final state particles multiplicity as well as
spin correlation when connecting production and decay processes.
\\
6) a new improved interactive numerical session which sums over the
subprocesses and connects production and decay processes in the GUI as
well as parallelizing over multiple cpus on a multicore machine.\\
Work in all these directions is in progress.
The \CalcHEP~team is open to additional sugegstions/requests from users.

\section*{Acknowledgements}

AB would like to thank the NExT Institute and SEPnet
for partial financial support. 
The work of  AB and AP was strongly supported by 
the Royal Society grant JP090598. 
AP was also supported by the Russian foundation for Basic Research, grants 10-02-01443-a
and 12-02-93108-CNRSL−a. 
The work of NC was supported in part by the US National
  Science Foundation under grant PHY-0354226, the LHC-TI initiative of
  the US National Science Foundation under grant PHY-0705682, the
  PITTsburgh Particle physics, Astrophysics, and Cosmology Center, and
  the US Department of Energy under grant DE- FG02-12ER41832.

A large part of the recent developments in \CalcHEP~were motivated by requests for the micrOMEGAs
package which uses \CalcHEP~as a generator of matrix elements. We thank
Genevieve Belanger and  Fawzi Boudjema  for helpful comments and suggestions in the development of
\CalcHEP.

\CalcHEP~is also know for its succesful usage of various SM extensions  mainly due to the
pioneering work of Andrey Semenov in the development of the LanHEP package
which allows to  realize rather complicated models in \CalcHEP~notation.
We are grateful to  Andrey Semenov for this job. 

{NC would like to thank Fabio Maltoni and Claude Duhr for their support
in writing a \CalcHEP~output for the FeynRules package.}

In 2009 AB and   AP  worked
together at the Galileo Galilei Institute for Theoretical Physics where
part of the \CalcHEP~development  was performed. We are grateful to GGI for  this opportunity.

We are especially thankful to Maksym Bondarenko, the main author of the HEPMDB project
which has turned out to be very important for various web applications of \CalcHEP.
  
We are grateful for many people who have tested \CalcHEP, especially to Lorenzo Basso, Alexandra Carvalho, Joydeep Chakrabortty, Qing-Hong Cao, Mikhail Chizhov, Asesh Datta, Mads Frandsen, Renato Guedes, K.C.  Kong, Tomas Lastovicka, Orlando Panella, Rui Santos, Chloe Papineau,  Marco Pruna, Patrik Svantesson, Florian Staub,  Daniel Stolarski, Riccardo Torre,  Izmail Turan and Alfonso Zerwekh.

\newpage
\bibliographystyle{elsarticle-num}
\newpage
\bibliography{bib}
\end{document}